\documentclass[12pt,titlepage,natbib209,twocolumn]{prop}

\usepackage[letterpaper,margin=1in]{geometry}
\usepackage{html,hyperref}
\usepackage[pdftex]{graphicx}
\usepackage[z-times,expert,alpha]{tmsfnts}
\usepackage{amssymb}

\def\Mvir{M_{\rm vir}}
\def\Rvir{R_{\rm vir}}
\def\M200{M_{\rm 200}}
\def\Vmax{V_{\rm max}}
\def\rmax{r_{\rm max}}
\def\msub{m_{\rm sub}}

\def\msixs{m_{0.6}}
\def\mthree{m_{0.3}}
\def\mthrees{m_{0.3}}

\def\Lsun{\, L_{\odot}}
\def\Msun{\, M_{\odot}}

\begin{document}

\title{Dark matter substructure and dwarf galactic satellites}
\author{Andrey Kravtsov\thanks{email:{\tt andrey@oddjob.uchicago.edu}}}

\date{{\it\small Dept. of Astronomy \& Astrophysics, Kavli Institute for Cosmological Physics,\\
 The University of Chicago, Chicago, IL 60637, USA\\[3mm]
review paper submitted to the special issue ``Dwarf Galaxy Cosmology'' of Advances
in Astronomy}\\[5mm]
\begin{minipage}[h]{6.5in}
{\bf Abstract.} A decade ago cosmological simulations of increasingly
higher resolution were used to demonstrate that virialized regions of
Cold Dark Matter (CDM) halos are filled with a multitude of dense,
gravitationally-bound clumps. These dark matter {\it subhalos\/} are
central regions of halos that survived strong gravitational tidal
forces and dynamical friction during the hierarchical sequence of
merging and accretion via which the CDM halos form. Comparisons with
observations revealed that there is a glaring discrepancy between
abundance of subhalos and luminous satellites of the Milky Way and
Andromeda as a function of their circular velocity or bound mass
within a fixed aperture. This large discrepancy, which became known as
the ``substructure'' or the ``missing satellites'' problem, begs for
an explanation. In this paper I review the progress made during the
last several years both in quantifying the problem and in exploring
possible scenarios in which it could be accommodated and explained in
the context of galaxy formation in the framework of the CDM paradigm
of structure formation. In particular, I show that the observed
luminosity function, radial distribution, and the remarkable similarity of 
the inner density profiles of luminous satellites can be 
understood within hierarchical CDM framework using a simple model
in which efficiency of star formation monotonically decreases with decreasing
virial mass satellites had before their accretion {\it without any actual
sharp galaxy formation threshold.} 
\end{minipage}
}
\maketitle

\section{Introduction}
\label{sec:intro}

In the hierarchical scenario of galaxy formation \cite{white_rees78},
theoretically rooted in the Cold Dark Matter (CDM) structure formation
model \cite{blumenthal_etal84}, galaxies form via cooling and
condensation of gas in dark matter halos, which grow via an
hierarchical sequence of mergers and accretion.  The density
perturbations in these models have amplitude that increases 
with decreasing scale down to $\sim 1$ comoving parsec or below \cite{loeb_zaldarriaga05}, 
with the smallest fluctuation scale defined by the specific properties of the
particles assumed to constitute the majority of the CDM.
Smaller perturbations thus collapse first and then grow and merge to
form larger and larger objects, with details of the evolution
determined by expansion history of the universe (i.e., by parameters
describing the background cosmological model) and by the shape of the
density fluctuation power spectrum \cite{white96}. 

\begin{figure*}[t]
\centerline{\includegraphics[width=0.3\linewidth]{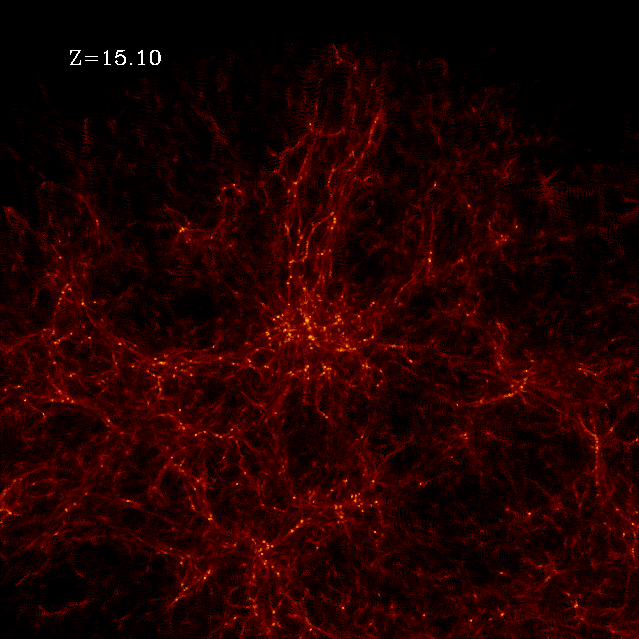}\hspace{0.5ex}\includegraphics[width=0.3\linewidth]{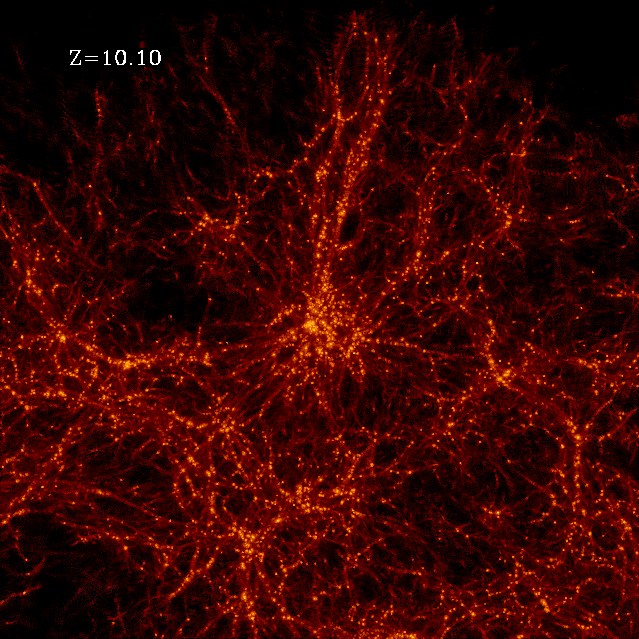}\hspace{0.5ex}\includegraphics[width=0.3\linewidth]{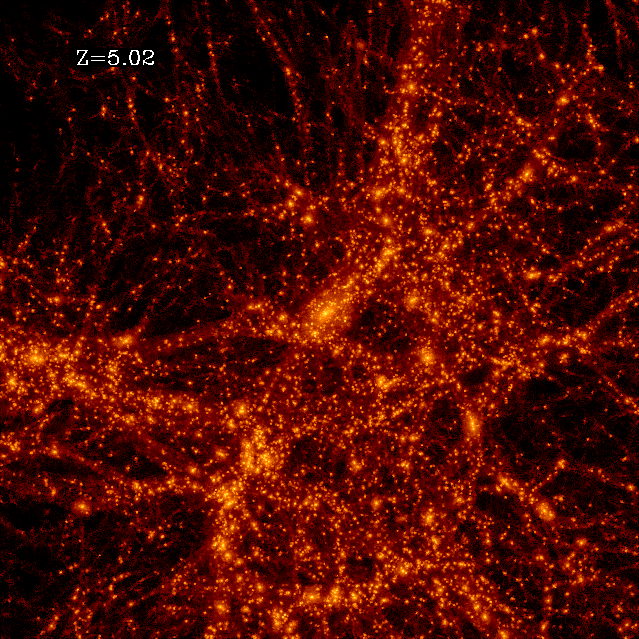}}\vspace{0.5ex}
\centerline{\includegraphics[width=0.3\linewidth]{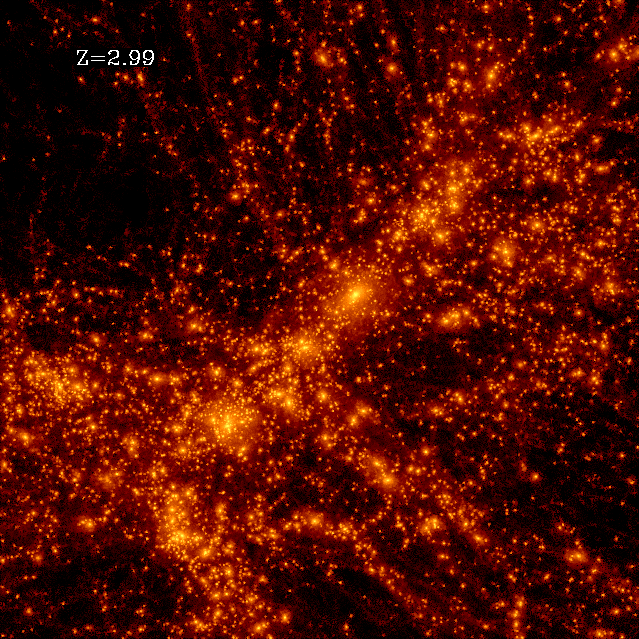}\hspace{0.5ex}\includegraphics[width=0.3\linewidth]{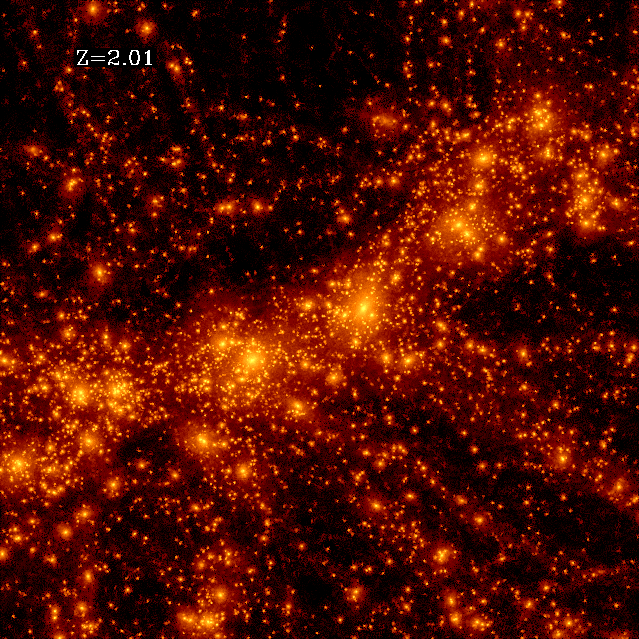}\hspace{0.5ex}\includegraphics[width=0.3\linewidth]{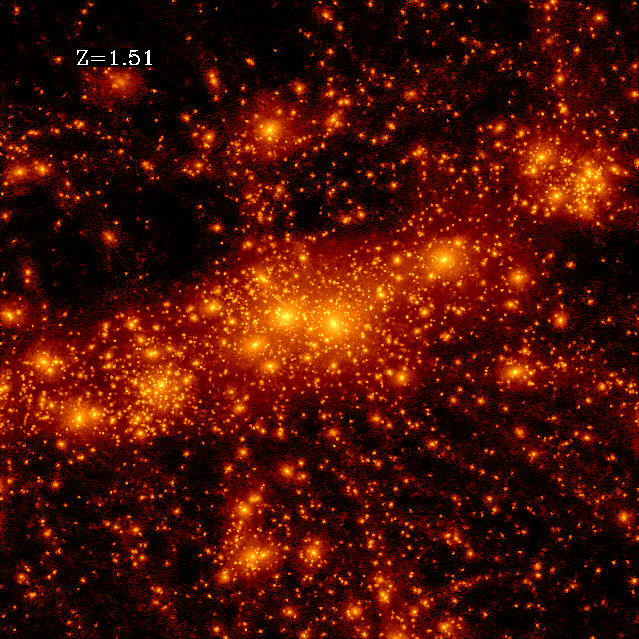}}\vspace{0.5ex}
\centerline{\includegraphics[width=0.3\linewidth]{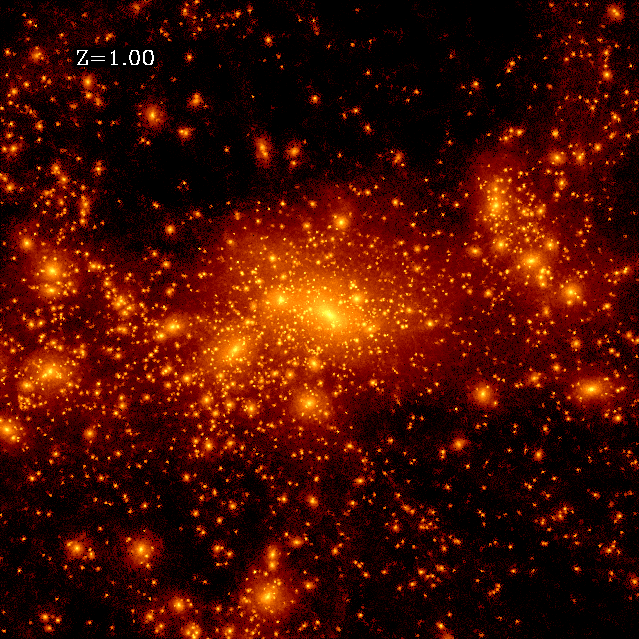}\hspace{0.5ex}\includegraphics[width=0.3\linewidth]{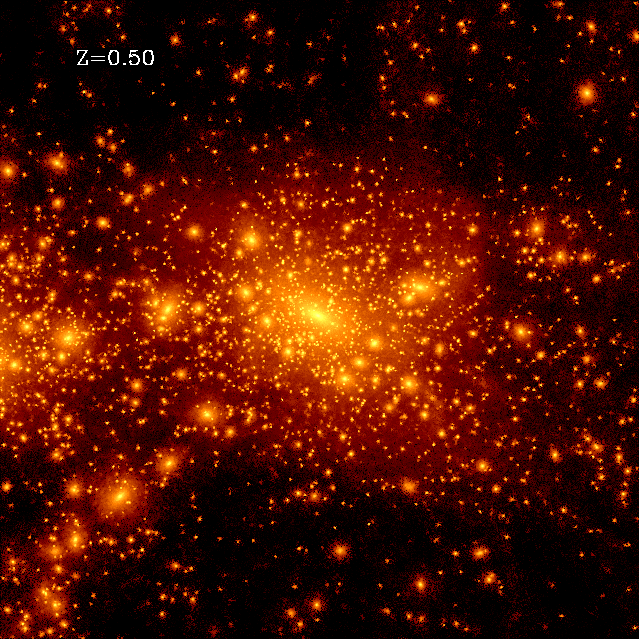}\hspace{0.5ex}\includegraphics[width=0.3\linewidth]{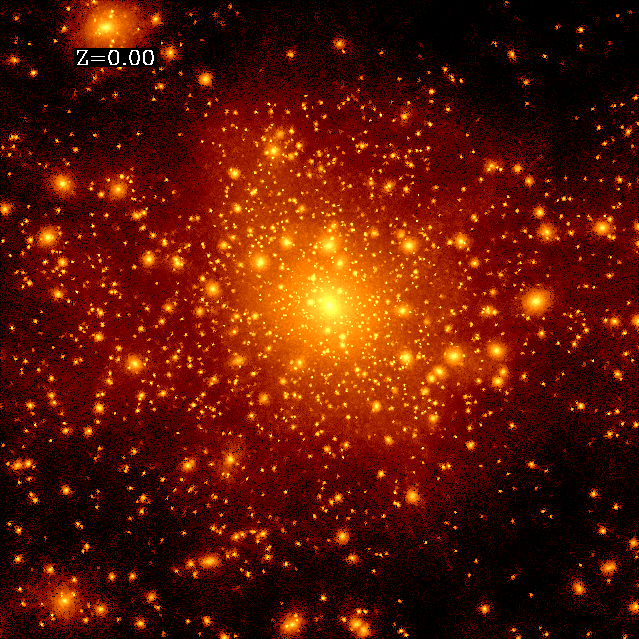}}

\caption{Formation of a Milky Way-sized dark matter halo in a cosmological simulation of flat
$\Lambda$CDM cosmology ($\Omega_{\rm m}=1-\Omega_{\Lambda}=0.3$, $h=0.7$, $\sigma_8=0.9$). 
The panels show an evolutionary sequence at nine redshifts (from left to right and from top to bottom) 
focusing on the most massive
progenitor of the main halo at each epoch (redshift of each epoch
is shown in the left upper corner). The rendering shows the
dark matter particles with intensity indicating the local matter
density on a logarithmic stretch. The build-up  of the halo proceeds
through a series of spectacular mergers, particularly frequent
in the early stages of evolution. Many of the merging clumps survive
until the present epoch ($z=0$) in the form of "substructure".
The size of the region shown is about 3 comoving Mpc at $z=15$, monotonically zooming in
to a scale of $\approx 1$ comoving Mpc across at $z=0$.}
\label{fig:hevol}
\end{figure*}

An example of such evolution in the flat $\Lambda$CDM model is
illustrated in Figure~\ref{fig:hevol}, which shows collapse of a
$\approx 10^{12}\Msun$ object. The figure shows that during the early stages of evolution the
matter that is incorporated into the final halo collapses into a large
number of relatively small clumps with a filamentary, web-like spatial
distribution. Further evolution, mediated by the competition between
gravity and expansion of space, is a sequence of accretion and
mergers that builds objects of progressively larger mass until the
single system is formed during the last several billion years of
evolution. The figure also shows that cores of some of the small
clumps that merge with and are incorporated into larger objects
survive\footnote{Survival of subhalos is not a trivial result and is
due to a relatively compact distribution of mass in CDM
halos. Ensuring their survival requires a rather large dynamic range
in spatial and mass resolution, which had not been achieved until the
late 1990s \cite{moore_etal96,tormen_etal98,klypin_etal99}.}  until later epochs and
are present in the form of {\it halo substructure\/} or {\it
subhalos}: small dense clumps within virialized regions of larger
halos.

The CDM model of structure formation is remarkably successful in
explaining a wide range of observations from temperature fluctuations
of the cosmic microwave background \cite{dunkley_etal09} to galaxy
clustering and its evolution \cite{kravtsov06} both qualitatively and,
in many cases, quantitatively. Nevertheless, many key details of the
model are still being developed \cite{birnboim_dekel03,keres_etal05}
and its testing is by no means complete.

One area of active investigation is testing predictions of the CDM
models at scales from a few kpc to tens of pc (i.e., the smallest
scales probed by observations of galaxies). In particular, there is
still tension between predictions of the central mass
distribution in galaxies \cite{deblok_etal08,deblok09} and sizes and angular momenta of galactic
disks and observational results \cite{mayer_etal08}. Notably, this tension has not
gone away during the past 10-15 years, even though both theoretical
models and observations have improved dramatically.

Another example of tension between CDM predictions and observations
that has been actively explored during the last decade is the fact
that satellite systems around galaxies of different luminosity are
qualitatively different, even though their dark matter halos are
expected to be approximately scaled down versions of each other
\cite{nfw97}, with their total mass as the scaling parameter. Faint
dwarf galaxies usually have no luminous satellites at all, Milky Way
and Andromeda have a few dozen, but clusters of galaxies often have
thousands of satellites around the brightest cluster galaxy. 

\begin{figure*}[t]
\vspace{-2.25cm}
\centerline{\includegraphics[width=1\linewidth]{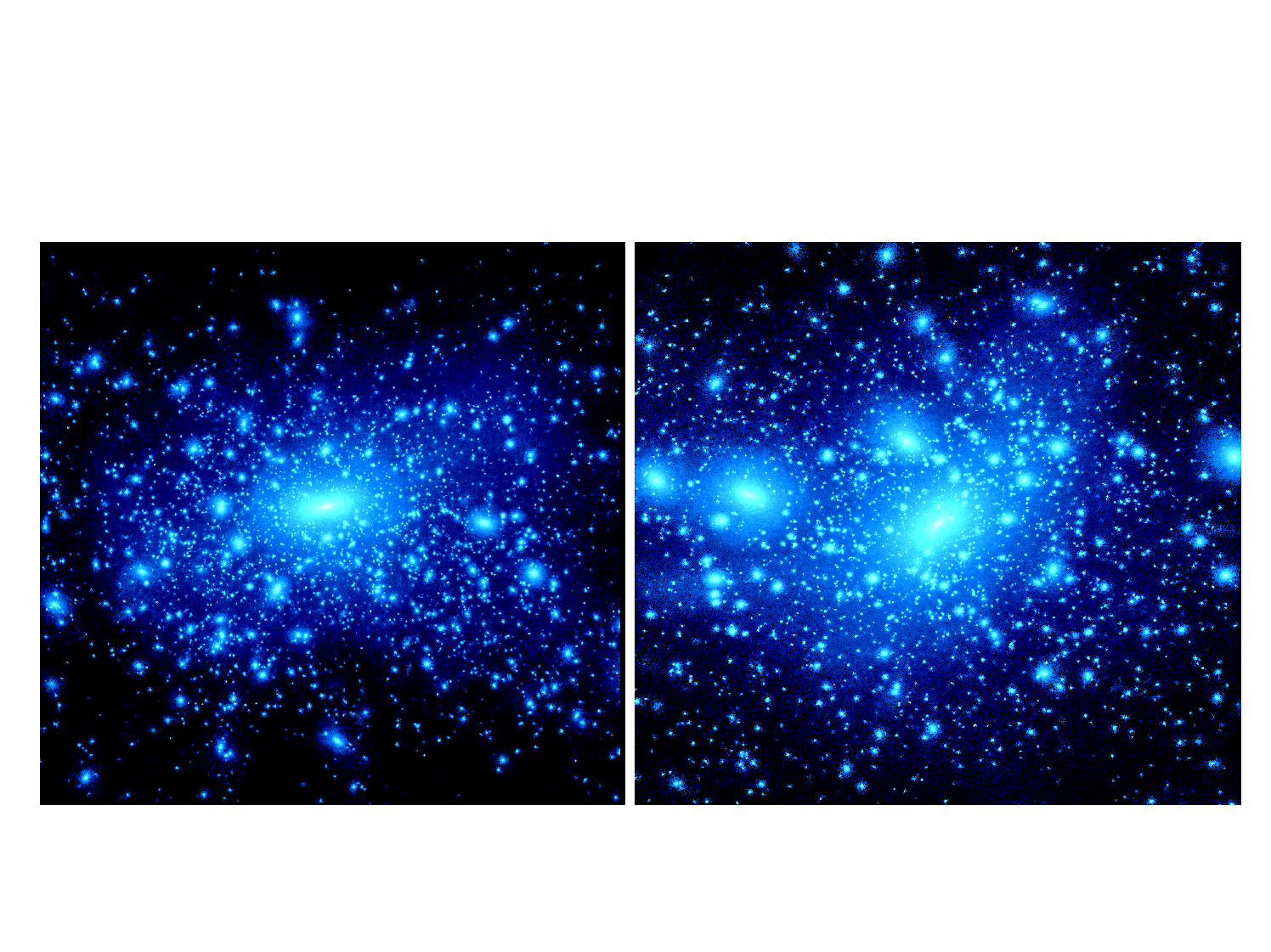}}
\vspace{-1.5cm}
\caption{Comparison of two $z=0$ halos of masses $3\times 10^{14}\Msun$ and $3\times 10^{12}\Msun$ formed in flat
$\Lambda$CDM cosmology ($\Omega_{\rm m}=1-\Omega_{\Lambda}=0.3$, $h=0.7$, $\sigma_8=0.9$). In each case 
the mass distribution around the center of the halo is shown to approximately two virial radii from the center
of each halo. Both objects
were resolved with similar number of particles and similar spatial
resolution relative to the virial radius of the halo in their respective simulations. I leave 
it as an exercise to the reader to guess the mass of the halo shown in each panel. 
}
\label{fig:selfsim}
\end{figure*}

The number of gravitationally-bound satellite subhalos at a fixed mass
relative to the mass of their host CDM halo, on the other hand, is
expected to be approximately the same
\cite{moore_etal99,kravtsov_etal04a,gao_etal04}.  This is illustrated
in Figure~\ref{fig:selfsim}, which shows distribution of dark matter
out to approximately two virial radii around the centers of two CDM
halos of masses different by two orders of magnitude.  It is clear
that it is not easy to tell the mass of the
halo by simply examining the overall mass distribution or by counting
the number of subhalos. This is a visual manifestation of 
approximate self-similarity of CDM halos of different mass. If we would
compare similar images of distribution of luminous matter around galaxies
and clusters, the difference would be striking. 

The manifestly different observed satellite populations around
galaxies of different luminosities and expected approximately
self-similar populations of satellite subhalos around halos of
different mass is known as the {\it substructure problem.}
\cite{kauffmann_etal93,klypin_etal99a,moore_etal99}.  In the case of
the best studied satellite systems of the Milky Way and Andromeda
galaxies, the discrepancy between the predicted abundance of
small-mass dark matter clumps and the number of observed luminous
satellites as a function of circular velocity (see \S~\ref{sec:quant})
has been also referred to as the {\it ``missing satellites
problem.''}\footnote{The name derived from the title ``Where are the
missing galactic satellites?'' of one of the papers originally pointing
out the discrepancy \cite{klypin_etal99a}.}  The main goal of this
paper is to review theoretical and observational progress in
quantifying and understanding the problem over the last decade.

\section{Quantifying the substructure and luminous satellite populations}
\label{sec:quant}

In order to connect theoretical predictions and observations on a
quantitative level, we need descriptive statistics to characterize
population of theoretical dark matter subhalos and observed luminous
satellites. Ideally, one would like theoretical models to be able to
predict properties of stellar populations hosted by dark matter halos
and subhalos and make comparisons using statistics involving directly
observable quantities, such as galaxy luminosities. In practice,
however, this is difficult as such predictions require modeling of
still rather uncertain processes shaping properties of galaxies during
their formation. In addition, the simulations can reach the highest
resolution in the regime when complicated and computationally costly
galaxy formation processes are not included and all of the matter in
the universe is modeled as a uniform collisionless and
dissipationless\footnote{That is the component that cannot dissipate the
kinetic energy it acquires during gravitational collapse and
accompanying gravitational interaction and relaxation processes.}
component. Such simulations thus give the most accurate knowledge of
the dark matter subhalo populations, but can only predict dynamical
subhalo properties such as the depth of their potential well or the
total mass of gravitationally bound material.  Therefore, in
comparisons between theoretical predictions and observations so far,
the most common strategy was to find a compromise quantity that can be
estimated both in dissipationless simulations and in observations.

\subsection{Quantifying the subhalo populations.} 

Starting with the first studies that made such comparisons using
results of numerical simulations \cite{klypin_etal99a,moore_etal99}
the quantity of choice was the {\it maximum circular velocity,}
defined as
\begin{equation}
\Vmax=\max\left(\frac{Gm(<r)}{r}\right)^{1/2},
\end{equation}
where $m(<r)=4\pi\int \rho(r)r^2dr$ is the spherically averaged total
mass profile about the center of the object. $\Vmax$ is a measure of
the depth of the potential (the potential energy of a self-gravitating
system is $W\propto\Vmax^2$) and can be fairly easily computed in a
cosmological simulation once the center of a subhalo is
determined.\footnote{The detailed description of the procedure of
identifying the centers of subhalos is beyond the scope of this paper,
but is nevertheless pertinent.  While many different algorithms are
used in the literature
\cite{ghigna_etal98,klypin_etal99,springel_etal01,diemand_etal04,gill_etal04,weller_etal05,knollmann_knebe09},
all algorithms boil down to the automated search for density peaks 
(most often in configuration space, but sometimes in the phase space)
field smoothed at a scale comparable to or smaller than the size of the
smallest subhalos in the simulations. Once the peaks are identified,
the gravitationally bound material around them is usually found by
iteratively removing the unbound particles.} The attractive feature 
of $\Vmax$ is that it is well defined and does not require estimate 
of a physical boundary of subhalos, which is often hard to determine. 
The price is that resolution required to get the $\Vmax$ correctly 
for a subhalo is higher, compared for example to the total bound mass
of subhalo, because 
$\Vmax$ is probing the mass distribution in the inner regions of halos. 

The total gravitationally bound mass of a subhalo, $\msub$, is less
sensitive to the resolution, but requires careful separation between
real subhalo particles and unbound particles from the diffuse halo of
the host dark matter halo. This can be quite difficult in the inner
regions of the host system where density of the background diffuse
halo is comparable to the internal density of subhalo or when two
subhalos overlap substantially. 

An alternative option is to define mass of a subhalo within a fixed
physical radius. For suitably chosen radius value, the mass can be
measured unambiguously both in simulations and in observations. 
We will discuss the measurement of the enclosed mass and comparisons between 
simulations and observations below in \S~\ref{sec:defining} and \S~ref{sec:models}
(see Figs.~\ref{fig:mf06} and \ref{fig:mf06model}). 

\begin{figure}[t]
\vspace{-2.25cm}
\centerline{\includegraphics[width=1.5\linewidth]{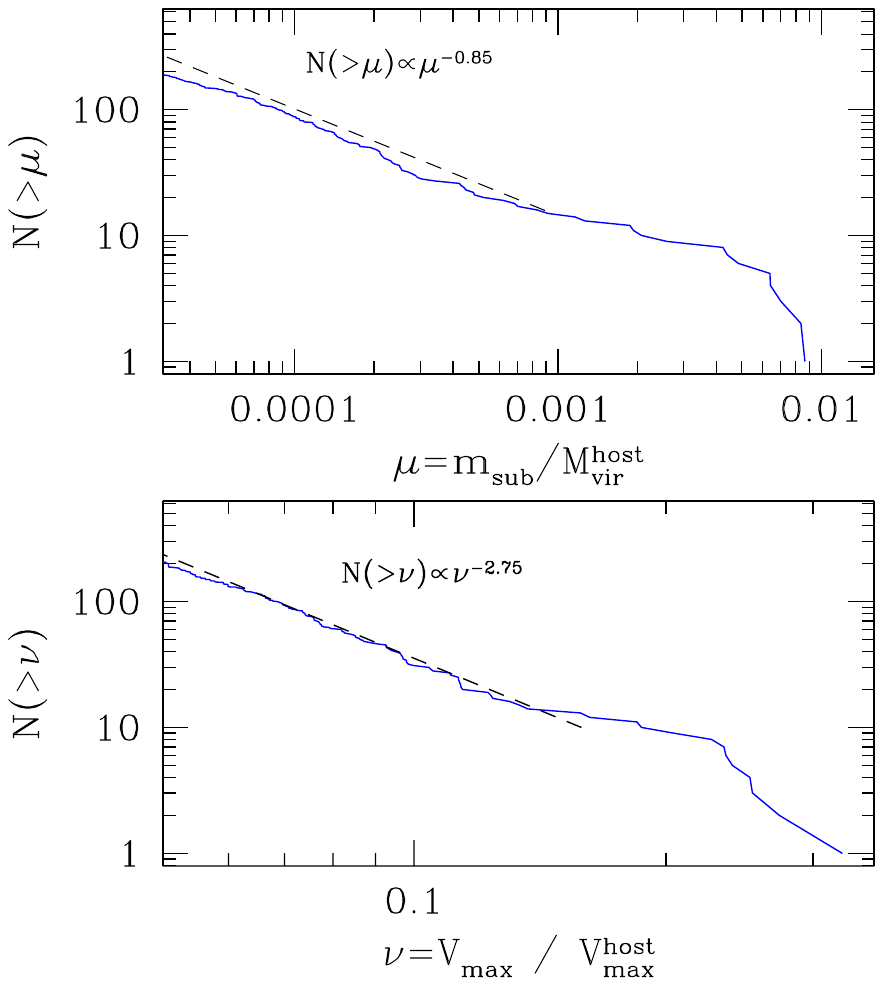}}
\vspace{-3cm}
\caption{Cumulative circular velocity and mass function of subhalos within
the virial radius $R_{337}=328$~kpc of a halo of virial mass $M_{337}=2\times 10^{12}\,\Msun$ at $z=0$.
The dashed lines show power laws with the slopes indicated in the legend of each panel.   }
\label{fig:vfmf}
\end{figure}

Figure~\ref{fig:vfmf} shows the cumulative circular velocity and mass
functions (CVF and CMF) of subhalos within the virial
radius\footnote{Defined as
$R_{\Delta}=(3M_{\Delta}/4\pi\Delta\bar{\rho})^{1/3}$, where
$\bar{\rho}$ is the mean matter density in the universe and
$\Delta=337$, corresponding to the $z=0$ virial overdensity suggested
by the spherical collapse model in the $\Lambda$CDM cosmology
\cite{lahav_etal91}.} of a simulated Milky Way sized halo, formation
of which was illustrated in Figure~\ref{fig:hevol}. Both cumulative
functions can be approximated by power laws over the ranges of circular velocity
and in units of $\Vmax$ and virial mass of the host:
$\nu\equiv V_{max}/\Vmax^{\rm host}\lesssim 0.1$ and $\mu\equiv
\msub/\Mvir^{\rm host}\lesssim 0.001$ with the slopes of $-3.7\div -4$
and $\approx -0.8\div -0.9$, respectively.  At large circular
velocities deviations from the power law can be significant due to
small numbers of subhalos.

The highest resolution simulations (to date) of the individual
MW-sized DM halos formed in the concordance $\Lambda$CDM cosmology
\cite{diemand_etal08,springel_etal08,stadel_etal09} show that the
power laws with the slopes in the range indicated above describe the
CVF and CMF down to $\mu\approx 10^{-7}$ and $\nu\approx
10^{-2}$. Note, however, that over a wider range of subhalo masses the
power law can be expected to change slowly reflecting the changing
slope of the rms fluctuations as a function of scale, which controls
the abundance of halos as a function of mass
\cite{jenkins_etal01,tinker_etal08}.

The amplitude of the mass and velocity functions is sensitive to the
normalization of the power spectrum on small scales
\cite{zentner_bullock03,diemand_etal08} and is thus sensitive to the
cosmological parameters that control the normalization (such as tilt
and normalization $\sigma_8$).  

For a given cosmology, the normalization of the CVF and CMF scales
approximately linearly with the host halo mass
\cite{kravtsov_etal04a}: $N(>\mu\vert M_{\rm h})\propto M_{\rm h}$.
The halo-to-halo scatter in the normalization of CVF and MCF 
for a fixed host halo virial mass is described by the Poisson
distribution \cite{kravtsov_etal04a}:
$\sigma_{N(>\mu)}=\sqrt{N(>\mu)}$. The fractional scatter is
therefore quite small for small $\mu$ and $\nu$ (large $N$).

The mass and circular velocity functions within a given radius
describe the overall abundance of subhalos of different mass, but not
their radial distribution. The latter depends rather sensitively on
how the subhalo samples are selected \cite{nagai_kravtsov05}. This is
because subhalos at different distances from their host halo center on
average experience different tidal mass loss, which affects different
subhalo properties by different amount. Subhalo mass is the most
affected quantity as large fraction of halo mass when it accretes
is relatively loosely bound and is usually lost quickly. Although
circular velocity is determined by the inner mass distribution in the inner mass
of subhalos, it is still affected by tidal stripping (albeit to a
lesser degree and slower than the total mass
\cite{kravtsov_etal04b}). 

The average mass loss experienced by subhalos increases with
decreasing distance to the center of the host halo
\cite{nagai_kravtsov05}. Therefore, selecting subhalos based on their
current bound mass or circular velocity biases the sample against
subhalos at smaller radii and results in the radial distribution much
less concentrated than the overall mass distribution of the host halo
\cite{ghigna_etal98,colin_etal99,diemand_etal04,gao_etal04b,kravtsov_etal04b,nagai_kravtsov05,maccio_etal06}.
Conversely, one can expect that if the selection of subhalos is made using
a quantity not affected by stripping, the bias should be smaller or even
disappear altogether.

\begin{figure}[t]
\vspace{-2.25cm}
\centerline{\includegraphics[width=1.5\linewidth]{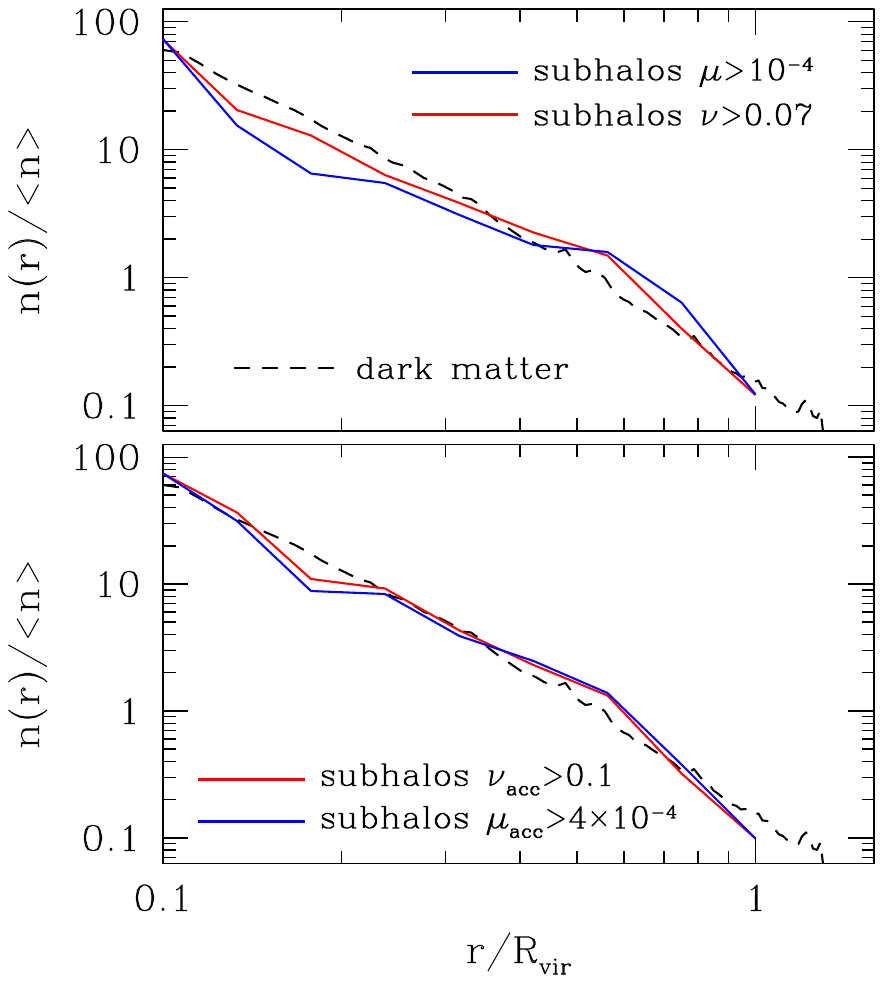}}
\vspace{-3cm}
\caption{Radial distribution of subhalos (solid lines) selected using
their different properties ($\Vmax$ and total mass --- solid red and
blue lines, respectively) compared to the matter density profile in
their host halo (dashed lines). The upper panel shows the profiles for
subhalos selected using circular velocity and bound mass a the current
epoch, while the lower shows radial distribution of subhalos selected
using the corresponding quantities {\it before} subhalo was accreted
unaffected by the subsequent tidal mass loss. Note that minimum
threshold values for subhalo selection, $\mu_{\rm min}$ and $\nu_{\rm
min}$, are different in the lower panel because for a typical mass
loss many subhalos with smaller circular velocities and masses at the
time of accretion fall below the completeness limit of the simulation
by $z=0$. }
\label{fig:rad}
\end{figure}

The top panel of Figure~\ref{fig:rad} shows the radial distribution of
subhalos (the same population as in Figure~\ref{fig:vfmf}) selected
using their current bound mass or circular velocity and density
profile of dark matter within a MW-sized sized host halo.  The figure
shows that the subhalo distribution is less radially concentrated
compared to the overall density profile because selection using
current subhalo properties affected by tidal evolution biases the
sample against the inner regions. The bottom panel shows the radial
distribution of subhalos in the same host halo but now selected using
circular velocity and mass the subhalos had {\it before accretion\/}
(which are of course not affected by the tides). In this case the
radial profile is very close to that of the dark matter
distribution. This dependence of the radial profile on the property
used for subhalo selection should be kept in mind when the observed
and predicted radial distributions are compared.  The latter are
selected based on their luminosity (i.e., the stellar mass), which may
be affected by tides much less than either the total bound mass or
circular velocity \cite{nagai_kravtsov05,penarrubia_etal08b}.

Finally, the spatial distribution of satellites is not completely 
spherically symmetric, but is triaxial, which reflects their accretion
along filaments and subsequent evolution in the triaxial potential of their
host halos \cite{zentner_etal05,libeskind_etal05}. 

\subsection{Quantifying populations of luminous galactic satellites.}

Although we currently know only a few dozen of nearby satellite
galaxies around the Milky Way and Andromeda, these galaxies span a
tremendous range of the stellar densities and luminosities. The two
brightest satellites of the Milky Way, the Large and Small Magellanic
Clouds (LMC and SMC), are easily visible by the naked eye in the
southern hemisphere and have therefore been known for many hundreds of
years, while the faintest satellites have been discovered only very
recently using sophisticated search algorithms and the vast data sets
of stellar photometry in the Sloan Digital Sky Survey and contain only
a few hundred stars \cite{belokurov_etal07,geha_etal09}.  Up until
late 1990s only a dozen dwarf galaxies were known to exist within 300
kpc of the Milky Way, with a similar number around the Andromeda
\cite{mateo98}. These galaxies have luminosities $L\gtrsim
10^5\,\Lsun$ and morphologies of the three types: 1) dwarf irregular
galaxies (dIrrs, e.g., LMC and SMC) --- low surface brightness
galaxies of irregular appearance which have substantial amount of gas
and thus exhibit continuing star formation, 2) dwarf spheroidal
galaxies (dSphs, e.g., Draco or Fornax) --- low surface brightness
galaxies with spheroidal distribution of stars and no (or very little)
ongoing star formation and 3) dwarf elliptical galaxies (dEs,
e.g. M32) --- high-surface brightness, low-luminosity ellipticals with
no gas and no current star formation. dSph and dE galaxies tend to be
located within 200 kpc of their host galaxies, while dIrr galaxies 
are distributed more uniformly. This tendency is called the ``morphological
segregation'' (\cite{mateo98}) and appears to exist in other
nearby groups of galaxies \cite{stierwalt_etal09}.
The properties of these
``classical'' dwarf galaxies are reviewed extensively by
\cite{mateo98} (see also recent study of scaling relations of dwarf
galaxies by \cite{woo_etal08}.

Despite the wide range of observed properties, all of the nearby
dwarfs share some common features in their star formation histories
(SFHs). The SFHs of all classical dwarfs is characterized by a rather
chaotically varying star formation rates. Most bright dwarfs form
stars throughout their evolution, although the majority of stars may
be formed in several main episodes spread over ten or more billion
years \cite{grebel_etal04,dolphin_etal05,orban_etal08,tolstoy_etal09}, and have at least some
fraction of stars that formed in the first two billion years of the
evolution of the universe.  In terms of their SFHs, the main
difference between the dwarf irregular and dwarf spheroidal galaxies
is in the presence or absence of star formation in the last
two billion years \cite{orban_etal08}. 

\begin{figure}[t]
\vspace{-2.25cm}
\centerline{\includegraphics[width=1.5\linewidth]{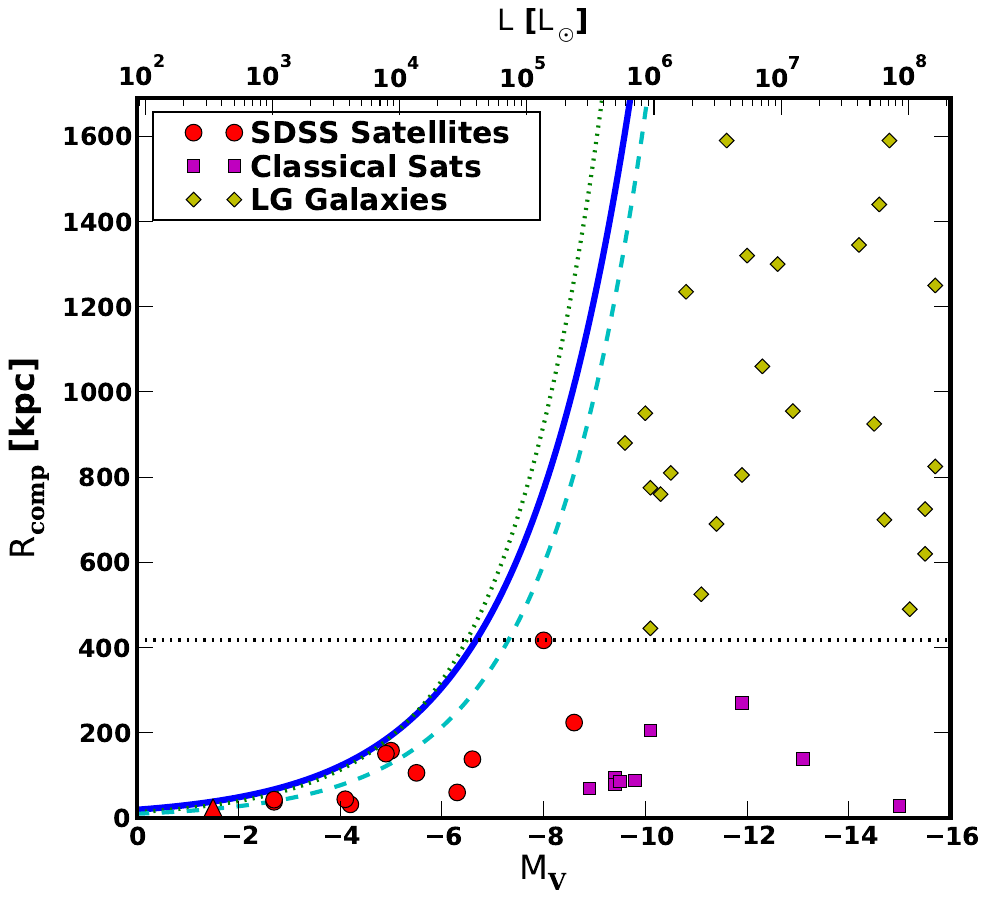}}
\vspace{-3cm}
\caption{The distance to which the current samples of dwarf satellite
galaxies around the Milky Way are complete as a function of galaxy
luminosity (absolute V-band magnitude on the bottom scale and physical
luminosity in units of solar luminosity at the top scale). Note that
the completeness distance of the faintest recently discovered
satellites is $\lesssim 50$~kpc. Adopted from
\protect\cite{tollerud_etal08}.  }
\label{fig:complete}
\end{figure}

\begin{figure}[t]
\vspace{-2.25cm}
\centerline{\includegraphics[width=1.5\linewidth]{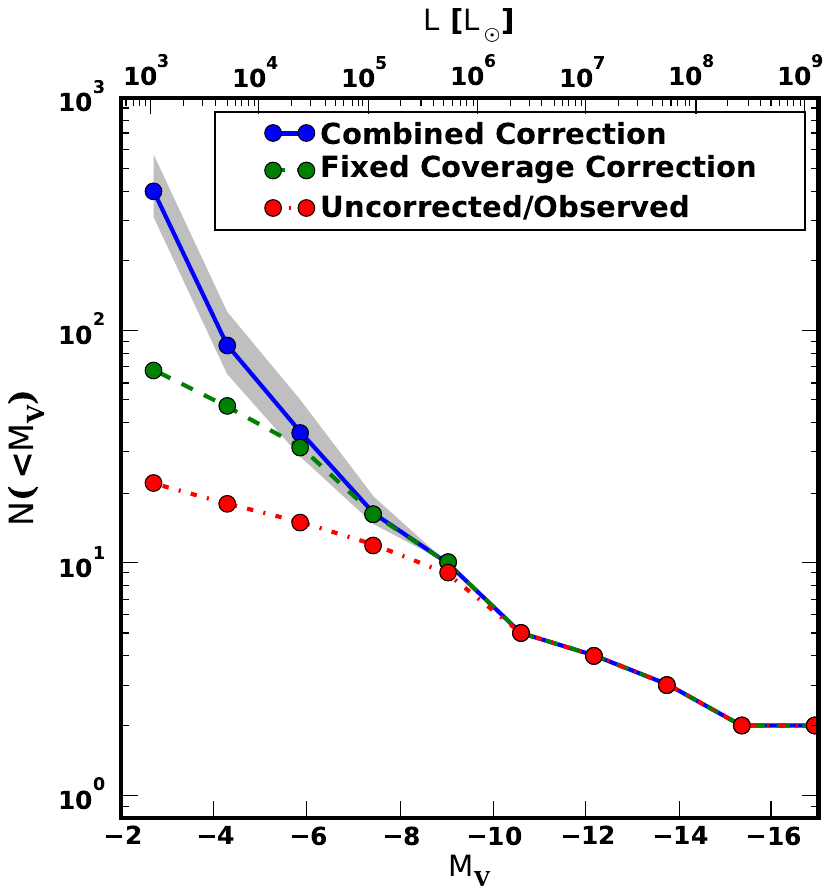}}
\vspace{-3cm}
\caption{The luminosity function of dwarf galaxies around the Milky
Way. The function includes all observed galaxies within 417 kpc of the
Milky Way. The red circles connected by the dotted line show the luminosity
function as observed, without any corrections for incompleteness. The green
circles connected by the dashed line show the observed luminosity function
corrected for limited coverage of surveys on the sky. The blue circles
connected by the solid line show luminosity function corrected
for the radial bias using radial distribution of subhalos from the Via
Lactea I simulation. Adopted from \protect\cite{tollerud_etal08}.  }
\label{fig:lf}
\end{figure}

The radial distribution of the classical satellites around the Milky
Way is rather compact. For the dwarf galaxies within 250 kpc of the
Milky Way, the median distance to the center of the Galaxy is $\approx
70$~kpc \cite{kravtsov_etal04b,willman_etal04}, while the predicted
median distance for subhalos is $\approx 120-140$~kpc
\cite{kravtsov_etal04b} (see Figure~\ref{fig:rdmodel} below). The
distribution of the satellites around the Andromeda galaxies is
consistent with that of the Milky Way satellites, but is less
accurately determined due to larger errors in distances. The spatial
distribution of satellites about the Milky Way and Andromeda is also
manifestly non-isotropic with the majority of the satellites found in
a flattened structure nearly perpendicular to the disk
\cite{lynden_bell82,majewski94,hartwick00,kroupa_etal05,metz_etal09}.

In 1994, a new faint galaxy was discovered in the direction toward
the center of the
Galaxy (in the Sagittarius constellation,
\cite{ibata_etal94}). The galaxy is similar to other nearby dwarf
spheroidal galaxies in its properties but is remarkably close to
the solar system (the distance of only $\approx 28$~kpc)  and
is in the process of being torn apart by the tidal interactions with
the Milky Way. This interaction has produced a spectacular tidal tail,
which has wrapped several times along the orbit of the Sagittarius
dwarf \cite{law_etal05}.

The discovery of this new satellite has alerted researchers in the
field to the possibility that other satellites may be lurking
undiscovered in our cosmic neighborhood. The advent of wide field
photometric surveys, such as the Sloan Digital Sky Survey and the
targeted surveys of regions around the Andromeda galaxy, and new
search techniques has resulted in dozens of new satellite galaxies
discovered during the last decade \cite{willman_etal05,belokurov_etal06,belokurov_etal07,zucker_etal06a,zucker_etal06b,irwin_etal07,koposov_etal07,walsh_etal07,geha_etal09} with many more
discoveries expected in the near future \cite{tollerud_etal08,willman09}. The majority of
the newly discovered galaxies are fainter than the ``classical
dwarfs'' known prior to 1998. Due to their extremely low luminosities
(as low as $\sim 1000\,\Lsun$ in the case of the Segue 1
\cite{geha_etal09}) they have collectively been referred to as the
``ultra-faint'' dwarfs. Such low luminosities (and implied stellar
masses) indicate an extreme mode of galaxy formation, in which the
total population of stars produced during galaxy evolution is smaller
than a star cluster formed in a single star formation event in
more luminous galaxies.

More practically, the extreme faintness of the majority of dwarf satellites
implies that we have a more or less complete census of them only
within the volume of $\sim 30-50$~kpc of the Milky Way
\cite{koposov_etal08,tollerud_etal08}.  Figure~\ref{fig:complete} shows the
distance to which the dwarfs of a given luminosity are complete in the
SDSS survey, in which the faintest new dwarfs have
been discovered. The figure shows that we have a good census of the
volume of the Local Group only for the relatively bright luminosities
of the ``classical'' satellites. At the fainter luminosities of
the ultra faint dwarfs, on the other hand, we can expect to find many more
systems at larger radii in the future deep wide area surveys. The exact
number we can expect to be discovered depends on their uncertain
radial distribution, but given the numbers of already discovered
dwarfs and our current knowledge of the radial distribution of brighter
satellites (and expected radial distribution of subhalos) we can reasonably expect that at least a hundred faint
satellites exist within 400 kpc of the Milky Way. This is illustrated in 
 Figure~\ref{fig:lf}, which shows the luminosity
function of the Milky Way satellites corrected for the 
volume not yet surveyed under different assumptions about radial distribution of the
satellites \cite{tollerud_etal08}.

The basis for considering these extremely faint stellar systems as
bona fide galaxies is the fact that, unlike star clusters, they are
dark matter dominated: i.e., the total mass within their stellar
extent is much larger than the stellar mass expected for old stellar
populations \cite{geha_etal09}. The total dynamical masses of these galaxies are derived
using kinematics of stars.\footnote{These faint dwarf spheroidal
galaxies do not have cold gas and therefore their mass profiles cannot be
measured using the gas rotation curve, as is commonly done for more
massive dIrr galaxies.} High resolution spectroscopy of the red giant
stars in the vicinity of each galaxy provides the radial velocities of
these stars.  The radial velocities can then be modeled using using
the Jeans equilibrium equations to derive the total mass profile
\cite{strigari_etal06,klim_etal07,wyse_gilmore08,penarrubia_etal08,walker_etal09,strigari09}. This
modeling requires certain assumptions about the unknown shape of the
stellar distribution and velocity distribution of stars, as well as
assumptions about the shape and radial profile of the dark matter
distribution. The resulting mass profile therefore has some
uncertainty associated with these assumptions
\cite{strigari_etal06,penarrubia_etal08,strigari09}.

Additionally, the ultra-faint dwarfs follow scaling relations
of the brighter classical satellites such as the luminosity-metallicity
relation \cite{kirby_etal08} and therefore seem to be the low luminosity
brethren within the family of dSph galaxies.

\section{Defining the substructure problem.}
\label{sec:defining}

As I noted above, comparison of theory and observations in terms of the
directly observable quantities such as luminosities is possible only
using a galaxy formation model. These models, although actively
explored
(\cite{bullock_etal00,somerville02,benson_etal02,kravtsov_etal04b,koposov_etal09,li_etal09,maccio_etal09,busha_etal09}, see also \S~\ref{sec:models})
are considerably more uncertain than the predictions of
dissipationless simulations on the properties of dark matter
subhalos. Given that observed dwarf satellites are very dark matter
dominated, the dissipative processes leading to formation of their
stellar component are expected to have a limited effect on
the distribution of the dynamically dominant dark matter. Fruitful
comparison between simulation predictions and observations is
therefore possible if a quantity related to the total mass profile can
be measured in the latter.

\begin{figure}[t]
\vspace{-1.75cm}
\centerline{\includegraphics[width=1.15\linewidth]{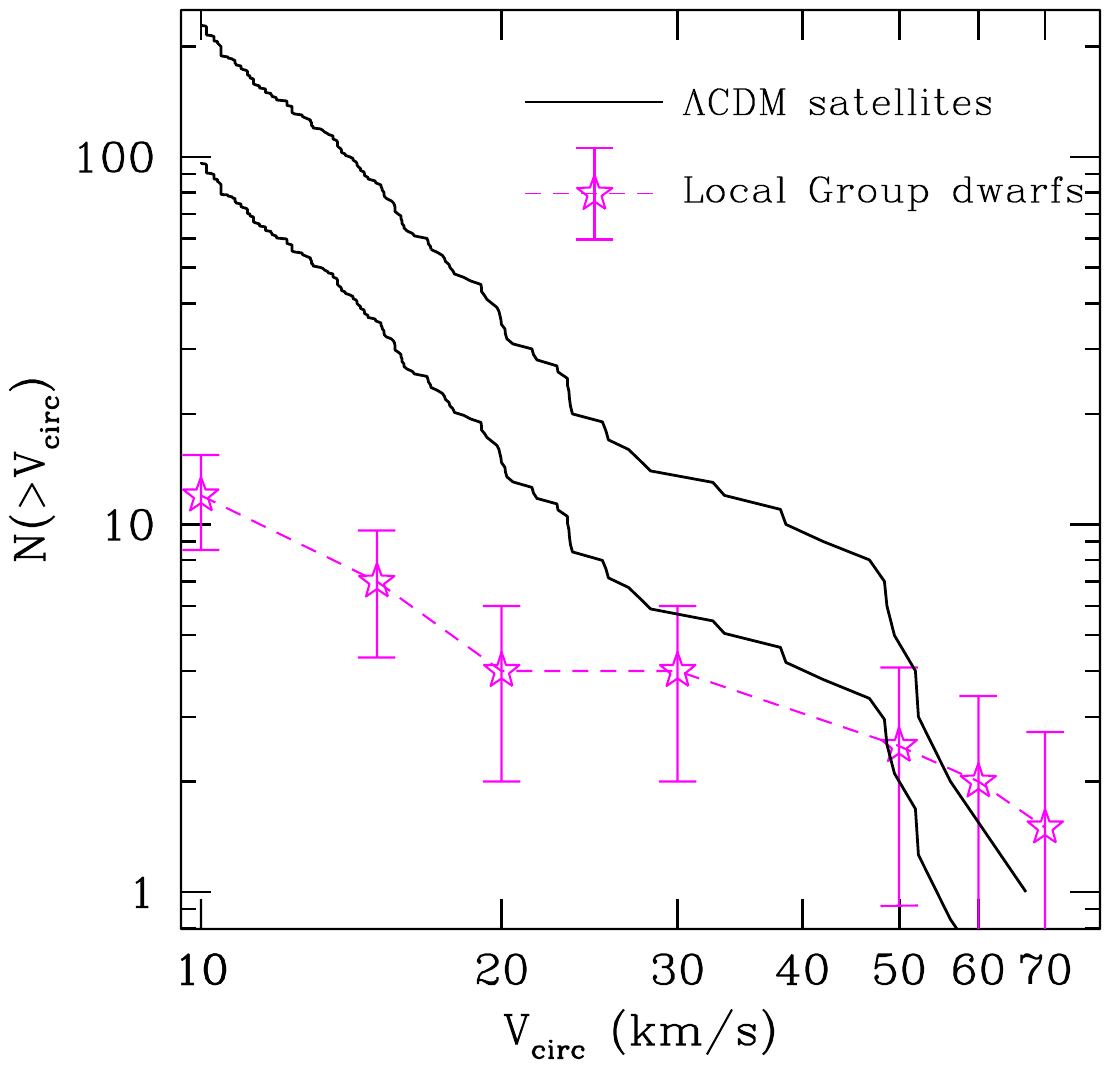}}
\vspace{-2.25cm}
\caption{Comparison of the cumulative circular velocity functions, $N(>\Vmax)$,
of subhalos and dwarf satellites of the Milky Way within the radius of 286 kpc (this
radius is chosen to match the maximum distance to observed satellites in the sample
and is smaller than the virial radius of the simulated halo, $R_{337}=326$~kpc). The subhalo VFs are
plotted for the host halos with max. circular velocities of $160$~km/s and $208$~km/s that should bracket the $\Vmax$ of the actual Milky Way halo. 
The VF for the observed satellites was constructed using circular
velocities estimated from the line-of-sight velocity dispersions as $\Vmax=\sqrt{3}\sigma_r$
(see discussion in the text for the uncertainties of this conversion).}
\label{fig:vflg}
\end{figure}

The first attempts at such comparisons
\cite{klypin_etal99,moore_etal99} assumed isotropy of the stellar
orbits and converted the line-of-sight velocity dispersion of stars in
dSph satellites, $\sigma_r$, to estimate their maximum circular
velocities as $\Vmax=\sqrt{3}\sigma_r.$ The admittedly over-simplistic
conversion was adopted simply due to a lack of well measured velocity
profiles and corresponding constraints on the mass distribution at the
time.  Figure~\ref{fig:vflg} shows such a comparison for the classical
satellites\footnote{I did not
include the new ultra-faint satellites in the comparison both because their $\Vmax$ values
are much more uncertain and because their total number within the virial radius
requires uncertain corrections from the currently observed number that probes
only the nearest few dozen kpc.  The velocity dispersions of the ultra-faint dwarfs are very similar
to each other ($\sim 5$~km/s) and they 
therefore formally have similar $\Vmax$ values according to this simple
conversion method (hence, they would all be ``bunched up'' at about the same $\Vmax\sim 9$~km/s value). The 
maximum circular velocity of the halos of these galaxies is expected to 
be reached at radii well beyond the stellar extent and its estimate
from the observed velocity dispersions requires substantial extrapolation
and assumptions about the density profile outside the radii probed by stars. 
The errors of the derived values of $\Vmax$ can therefore be 
quite substantial \cite{strigari_etal06,strigari_etal08}. I will compare the
predicted luminosity function of the luminous satellites using a simple galaxy
formation model in \S~\ref{sec:models} (see Fig.~\ref{fig:lfmodel}).}
 of the Milky Way and subhalo populations in Milky Way-sized
halos formed in the concordance $\Lambda$CDM cosmology. 

The observed velocity function is compared to the predicted VF of dark
matter subhalos within a 286 kpc radius of Milky Way-sized host halos.
In the literature, ``Milky Way-sized'' is often used to imply a total
virial mass of $\Mvir\approx 10^{12}\Msun$ and maximum circular
velocity of $\Vmax\approx 200$~km/s. However, there is some
uncertainty in these numbers. Therefore the figure shows the VFs for
the host halos with $\Vmax=208$~km/s and $160$~km/s. The former is
measured directly in a simulation of the halo of that circular
velocity, while the latter VF was rescaled as $N(>\Vmax)\propto M_{\rm
vir,1}/M_{\rm vir,2}=(V_{\rm max,1}/V_{\rm max,2})^{3.3}$, using
scaling measured statistically in the simulations
\cite{kravtsov_etal04a}.

The simple conversion of $\sigma_r$ to $\Vmax$ has justly been
criticized as too simplistic \cite{stoehr_etal02}.  Indeed, the
conversion factor $\eta\equiv\Vmax/\sigma_r$ requires a good knowledge
of mass profile from small radii to the radius $r_{\rm max}$. The mass
profile derived from the Jeans equation has errors associated with
uncertainties in the anisotropy of stellar orbits, as well as with
uncertainties of spatial distribution of stellar system and/or its
dynamical state \cite{strigari_etal07,klim_etal07}.  Most
importantly, the mass profile is only directly constrained within the
radius where stellar velocities are measured, $r_{\ast}$. If this
radius is smaller than $r_{\rm max}$, conversion factor $\eta$ depends
on the form of the density profile assumed for extrapolation. The
uncertainties of the derived mass profile within the stellar extent will
of course also be magnified increasingly with increasing
$r_{\max}/r_{\ast}$ ratio \cite{strigari_etal07}.

Thus, for example, Stoehr et al. \cite{stoehr_etal02} have argued that
the conversion factor can be quite large ($\eta\gtrsim 2-4$) if the density profile is
shallow in the inner regions probed by the stars. Such large conversion 
factor would shift the low $\Vmax$ points in Figure~\ref{fig:vflg} to the right closer to the
subhalo VF \cite{stoehr_etal02,hayashi_etal03,penarrubia_etal08} and 
would imply that there is a sharp drop in the luminosity function of satellites
below a certain threshold circular velocity ($\Vmax\approx 30$~km/s).
High-resolution simulations of individual satellites, on the other hand, have demonstrated
that the CDM satellites retain their {\it cuspy} inner density profiles,
even as they undergo significant tidal stripping \cite{kazantzidis_etal04}.
This implies that $\eta$ is likely not as large as advocated by Stoehr et al.
The main uncertainty in its actual value for a specific satellite is then 
due to the uncertainty in the density profile and ratio $r_{\rm max}/r_{\ast}$. 

Recently, Pe\~narrubia et al. \cite{penarrubia_etal08} combined the
measurements of stellar surface density and $\sigma_r(R)$ profiles to
estimate $\Vmax$ values for individual observed MW satellites under
assumption that their stellar systems are embedded into NFW dark
matter potentials. Such procedure by itself does not produce a
reliable estimate of $\Vmax$, because NFW potentials with a wide range
of $\Vmax$ can fit the observed stellar profiles. To break the
degeneracy Pe\~narrubia et al.  have used the relation between $\Vmax$
and $r_{\rm max}$ expected in the concordance $\Lambda$CDM
cosmology. They showed that this results in estimates of $\Vmax$ which
imply $\eta\approx 2-3$ (see their Fig.~4).  This estimate does not
take into account effects of tidal stripping on the evolution of the
$\rmax-\Vmax$ relation.  Typically, subhalos located
in the inner regions of the halo are expected to have lost $\sim 60-90\%$
of their initial mass by $z=0$ due to tidal stripping
\cite{kravtsov_etal04b,nagai_kravtsov05,giocoli_etal08}. For such
mass loss $\Vmax$ changes only by $\approx
20-30\%$ \cite{kravtsov_etal04b,penarrubia_etal08} but $\rmax$ should
change by a factor of $\approx 2-3$
\cite{kazantzidis_etal04,penarrubia_etal08b}.

In a subsequent study, Pe\~narrubia et al. \cite{penarrubia_etal08b} used
controlled simulations of subhalo evolution to  argue that tidal 
stripping does not significantly affect their inferred conversion factor $\eta$ (see
their Fig.~9). This conclusion, however, was drawn based on the systems
in which both stellar system and DM halo were significantly stripped. In such
system $r_{\rm max}$ is close to the stellar radius and $\sigma_0$ and 
$\Vmax$ evolve in sync. For systems with more realistic mass loss and with
stars deeply embedded within $r_{\rm max}$, however, 
stellar system (and $\sigma_0$) may not be affected, while $\Vmax$
can evolve significantly. For such systems the method of \cite{penarrubia_etal08} 
will lead to a significant overestimate of $\eta$. Indeed, the systems in Fig.~9 of \cite{penarrubia_etal08b}
for which the method overestimates $\eta$ (by a factor of $\approx 1.4$) the most are the systems
with moderate total mass loss and least affected stellar systems. Note
that even these systems have likely experienced more tidal loss than most of the 
real dSph satellites.

Another factor in estimates of $\Vmax$ is anisotropy of stellar
velocities in dSph (e.g., \cite{kazantzidis_etal04}).  For example,
recent analysis of observed velocity dispersion profiles of
``classical'' dSph by \cite{walker_etal09}, in which the anisotropy of
stellar orbits was treated as a free parameters, results in estimates
of $\Vmax$ of their host subhalos in the range $\sim 10-25$~km/s,
smaller than would be suggested if correction factor was $\eta\approx
2-3$.

Regardless of the actual conversion factor value, however, it is clear
that it cannot change the main difference between the observed and
predicted VFs -- the large difference in their slope -- unless $\eta$
strongly depends on $\Vmax$ (there is no observational evidence for
this so far).

\begin{figure}[t]
\vspace{-1.75cm}
\centerline{\includegraphics[width=1.2\linewidth]{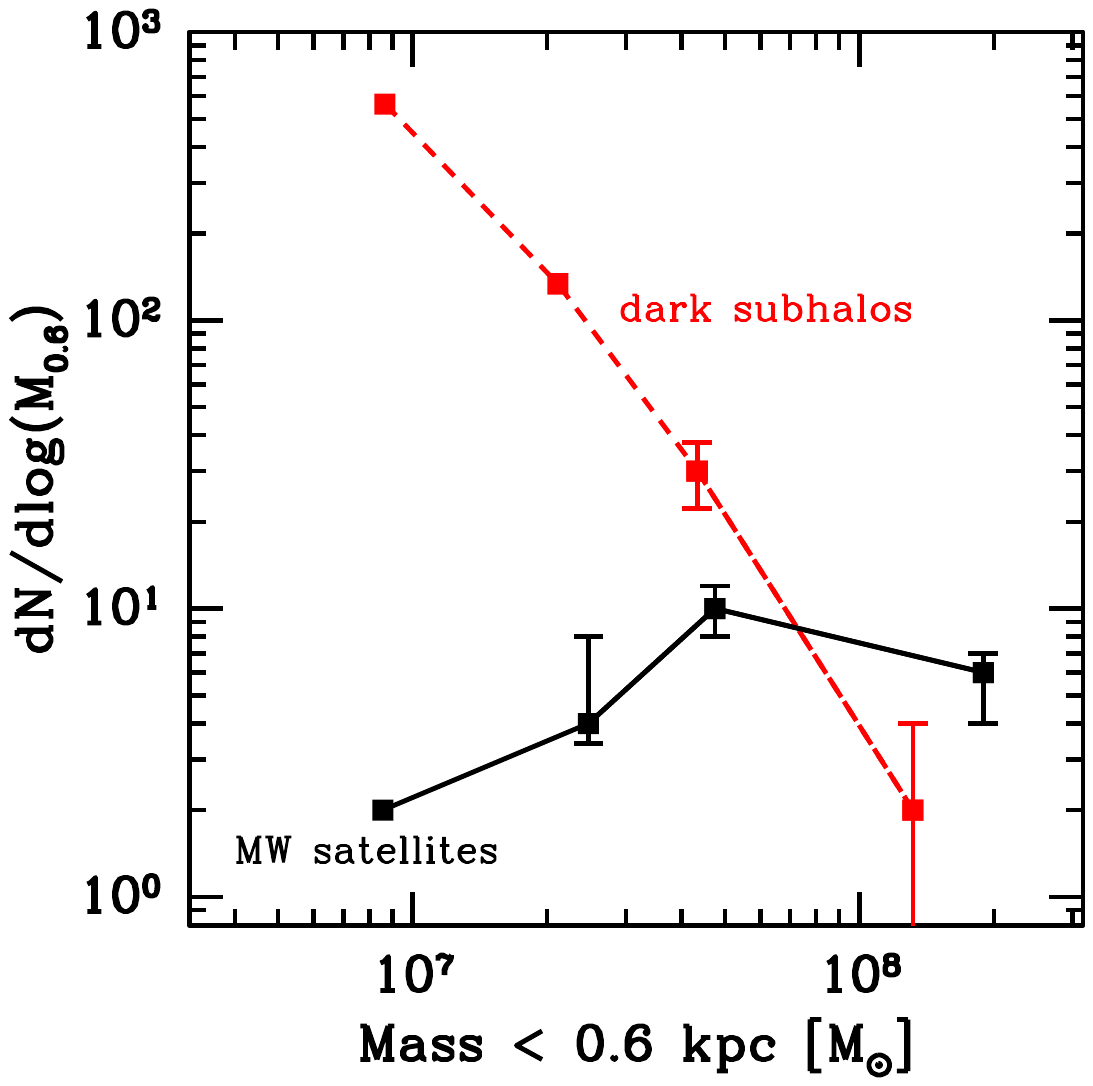}}
\vspace{-2.25cm}
\caption{The mass function of dwarf satellites of the Milky Way, where
masses of subhalos and observed satellites are measured within a fixed 
physical radius of $0.6$~kpc. Adopted from \protect\cite{strigari_etal07}.}
\label{fig:mf06}
\end{figure}

Another promising approach is to abandon attempts to derive $\Vmax$
altogether and to measure instead the observed mass within the radius
where the uncertainty of the measured mass profile is minimal. Such
radius is close to the stellar extent of observed galaxies
\cite{strigari_etal07,strigari_etal08,penarrubia_etal08,strigari09}.
Figure~\ref{fig:mf06}, adopted from \cite{strigari_etal07}, shows
comparison of the mass functions of subhalos in the Via Lactea I
simulation \cite{diemand_etal07} and observed satellites of the Milky
Way, where masses are measured within a fixed physical radius of
600~pc (see also discussion in \S~\ref{sec:models} and
Fig.~\ref{fig:mf06model}). The figure shows that the simulated and
observed mass functions are different, the conclusion similar to that
derived from the comparison of the circular velocity functions.

Thus, the discrepancy that is clearly seen in the comparison 
of circular velocity functions, measured with more uncertainty 
in observations, persists if the comparison is done using a much better measured
quantity. Unfortunately, the stellar distribution in most of the 
newly discovered ultra-faint dwarf galaxies does not extend out to 600~pc radius
and $m(r<0.6\,{\rm kpc})\equiv\msixs$ therefore cannot be measured as reliably for these faint
systems as for the classical dwarfs. Similar comparisons have to be carried out
using masses within smaller radii. This puts stringent requirements on the 
resolution of the simulations, as they need to reliably predict mass distribution
of subhalos within a few hundred parsec radius. Such high-resolution simulations
are now available, however \cite{diemand_etal08,springel_etal08,stadel_etal09}.  

We can draw two main conclusions from the comparisons of the circular
velocity functions and the more reliable $\msixs$ mass
functions of subhalos and observed satellites presented in the
previous sections, even taking into account existing uncertainties in
deriving circular velocities and the total dynamical masses for the
observed satellites.  First, the predicted abundance of the most
luminous satellites is in reasonable agreement with the data, even
though the statistics are small. Most MW-sized halos simulated 
in the concordance $\Lambda$CDM cosmology have 1-2 LMC sized ($\Vmax\approx 60-70$~km/s)
subhalos within their virial radius.

This is not a trivial fact because
the abundance of the most massive satellites is determined by a subtle
interplay between the accretion rate of systems of corresponding
circular velocity and their disruption by the combined effects of the
dynamical friction and tidal stripping
\cite{zentner_etal05b}. Dynamical friction causes satellites to sink to
the center at a rate which depends on the mass and orbital parameters of 
satellite orbit. Orbital parameters, in turn, depend on the cosmological environment
of the accreting host halo and are mediated by the tidal stripping
which reduces satellite mass as it sinks, thereby rendering
dynamical friction less efficient
\cite{colpi_etal99,hashimoto_etal03,boylankolchin_etal08,jiang_etal08}.
The fact that the concordance $\Lambda$CDM model makes an {\it ab initio}
prediction that the number of massive satellites that can host luminous
dwarfs is comparable to observations can therefore be viewed as a success
of the model. 

Second, the {\it slopes\/} of both the circular velocity function and the $\msixs$ mass
function are different in simulations and observations. This implies that 
we cannot simply match all of the luminous satellites to the subhalos
with the largest $\Vmax$ and $\msixs$, as was sometimes advocated \cite{stoehr_etal02,hayashi_etal03}.
The $\msixs$ mass function comparison, in particular, indicates that there
should be some subhalos with the $\msixs\sim 10^7\Msun$ that do not host
the luminous galaxies, and some that do. As I discuss in \S~\ref{sec:gfphysics}, 
this has a strong implication for the physical interpretation of the 
difference in terms of galaxy formation scenarios. 

In summary, {\it the substructure problem can be stated as
the discrepancy in the slopes of the circular velocity and $\msixs$ mass
functions inferred for observed satellites of the Milky Way and 
the slopes of these functions predicted for dark matter subhalos in
the MW-sized host halos formed in the concordance $\Lambda$CDM
cosmology.}

I believe that stated this way the problem is well-defined.  Defining
the problem in terms of the difference in the actual number of
satellites and subhalos is confusing at best, as both numbers are 
fairly strong functions of subhalo mass or stellar
luminosity. Thus, for example, even though the discovery of the
ultra-faint dwarfs implies the possible existence of hundreds of them in the halo of the Milky Way [49,75] (this fact has been used to argue that the substructure problem has been ``alleviated''), 
the most recent simulations show that more than 100,000 subhalos of
mass $\msub>10^5\Msun$ should exist in the Milky Way
\cite{diemand_etal08,springel_etal08}.\footnote{Indeed, it is not obvious
{\it a priori} that subhalos of mass $10^5-10^6\Msun$ are too small to
host luminous stellar systems of stellar mass $M_{\ast}\sim                                                                                                          10^3-10^4\, \Msun$ \cite{naoz_etal09} --- the stellar masses corresponding to the
luminosities of the faintest recently discovered dwarfs. After all, the
halos of this mass are expected to be hosting formation of the very first
stars \cite{abel_etal02}.}
 The substructure problem stated
in the actual {\it numbers} of satellites is therefore alive and well
and has not been alleviated in the least. 

I would like to close this section by a brief discussion of the
comparison of spatial distribution of observed satellites and
subhalos.  As I noted above, the radial distribution
of the observed satellites of the Milky Way is more compact than the
radial distribution of subhalos selected using their present day mass
or circular velocity \cite{kravtsov_etal04b,willman_etal04,taylor_etal05}. In
addition, the observed satellites are distributed in a quite
flattened structure with its plane almost perpendicular to the disk of
the Milky Way
\cite{lynden_bell82,majewski94,hartwick00,kroupa_etal05,metz_etal09}.
Although the spatial distribution of {\it all} subhalos is expected to be
anisotropic, reflecting the anisotropy of their accretion directions
along filaments \cite{zentner_etal05,libeskind_etal05} and, possibly,
the fact that some of the satellites could have been accreted as part
of the same group of galaxies \cite{donghia_lake08}, it is not as
strong as the anisotropy of the observed Milky Way satellites.

Thus, both the radial distribution and anisotropy of the observed
satellites do not match the overall distribution of subhalos in CDM
halos. This is likely another side of the same substructure problem
coin and the overall spatial distribution of observed satellites needs
to be explained together with the differences in the circular velocity
function. I will review a few possible explanations for the substructure
problem and differences in the spatial distribution in the next section.

\section{Possible solutions}
\label{sec:solutions}

\subsection{Modifications to the CDM model.}

One possible way to account for the differences
of the predicted and observed circular velocity and $\msixs$ mass functions is to assume that
$\Lambda$CDM model is incorrect on the small scales probed by the
dwarf galactic satellites. Indeed, the abundance of satellites 
is sensitive to the amplitude of the power spectrum on the scales
corresponding to the total mass of their host halos. For 
a halo of mass $M$ the comoving scale of fluctuations that seed their formation is 
\begin{eqnarray}
d&=&2R=2\left(\frac{3M}{4\pi\Omega_{\rm m0}\rho_{\rm crit0}}\right)^{1/3}\\\nonumber
&=&360.4\,{\rm kpc}\left(\frac{M}{10^9\,\Msun}\frac{0.3}{\Omega_{\rm m}}\right)^{1/3}\left(\frac{H_0}{70}\right)^{-2/3},
\end{eqnarray}
where $\Omega_{\rm m}$ is the present-day total matter density in
units of the present-day critical density, $\rho_{\rm crit0}\equiv
3H_0^2/8\pi G$ and $H_0$ is the current Hubble constant in units of
km/s/Mpc. 

If the amplitude of density fluctuations on such scales is
considerably suppressed compared to the concordance $\Lambda$CDM model
used in most simulations, the abundance of subhalos can then also be
suppressed. Such suppression can be achieved either by suitably
varying parameters controlling the amplitude of the small-scale power
spectrum within the $\Lambda$CDM model itself
\cite{zentner_bullock03}, such as the overall normalization of the
power spectrum or its large-scale tilt, or by switching to models in
which the amplitude at small scales is suppressed, such as the warm
dark matter (WDM) structure formation scenarios
\cite{kamionkowski_liddle00,colin_etal00,bode_etal01,zentner_bullock03,colin_etal08}. In these models the abundance of satellites is suppressed both because
fewer halos of dwarf mass form in the first place (due to smaller
initial amplitude of fluctuations) and because halos that do form have
a less concentrated internal mass distribution, which makes them more
susceptible to tidal disruption after they accrete onto their host
halo.  Models in which dark matter was assumed to be self-interacting,
a property that can lead to DM evaporation, have also been proposed and discussed
\cite{spergel_steinhardt00,moore_etal00}, but these models both run
into contradiction with other observational properties of galaxies and
clusters
\cite{kochanek_white00,yoshida_etal00,miralda00,gnedin_ostriker01} and
are now strongly disfavored by observational evidence indicating that
dark matter self-interaction is weak \cite{clowe_etal06}.

The problem, however, is more subtle than simply suppressing the
number of satellites. As discussed above, differences exist between
observed and predicted slopes of the circular velocity and mass.
 The slope is controlled by
the slope of the primordial fluctuation spectrum around the scale
corresponding to the masses of satellite halos and structural
properties of the forming halos.  It has not yet been demonstrated
convincingly whether {\it both\/} the circular velocity and the $\msixs$ mass
functions can be reproduced in any of the models alternative to
CDM. In fact, recent measurements of mass distribution in the central
regions of observed satellites put stringent constraints on the phase
space density and ``warmness'' of dark matter \cite{strigari_etal06}.
At the same time, measurements of the small-scale density power spectrum
of the Lyman $\alpha$ forest indicate that fluctuations with the amplitude 
expected in the $\Lambda$CDM model at the scales that control 
the abundance of dwarf mass halos are indeed present in the primordial
spectrum \cite{abazajian06,seljak_etal06}.

While the inner density distribution in observed satellites may still
be affected by dark matter warmness in the allowed range of parameter
space \cite{wyse_gilmore08} (see, however, \cite{walker_etal09}), the
models with such parameters would not suppress the overall abundance
of satellites considerably. In fact, observations of flux ratios in
the multiple image radio lenses appear to {\it require\/} the amount
of substructure which is even larger than what is typically found in
the CDM halos
\cite{mao_schneider98,dalal_kochanek02,kochanek_dalal04}, 
disfavoring models with strongly suppressed abundance of small mass
subhalos.

There is thus no compelling reason yet to think that the observed
properties of galactic satellite populations are more naturally
reproduced in these models.  In the subsequent discussion, I will
therefore use Occam's razor and focus on the possible explanations
of the differences between observed satellites and subhalos in
simulations within the $\Lambda$CDM model. The prime suspect in
producing the discrepancy is the still quite uncertain physics of galaxy
formation.  After all a similar problem exists for objects of larger masses and
luminosities if we compare the slope of the luminosity function and
the halo mass function \cite{gonzalez_etal00,baldry_etal08} or the
predicted and observed abundance of galaxies in the nearby low density
``field'' regions \cite{tikhonov_klypin09}.

\subsection{Physics of galaxy formation.}
\label{sec:gfphysics}

Several plausible physical processes can suppress gas accretion and
star formation in dwarf dark matter halos.  The cosmological UV
background, which reionized the Universe at $z\gtrsim 6$, heats the
intergalactic gas and establishes a characteristic time-dependent
minimum mass for halos that can accrete gas
\cite{thoul_weinberg96,quinn_etal96,gnedin_hui98,kitayama_ikeuchi00,gnedin00,dijkstra_etal04,hoeft_etal06,okamoto_etal08}.
The gas in the low-mass halos may be photoevaporated after
reionization \cite{barkana_loeb99,shaviv_dekel03,shapiro_etal04} or
blown away by the first generation of supernovae
\cite{dekel_silk86,maclow_ferrara99,dekel_woo03,mashchenko_etal08} (see, however,
\cite{marcolini_etal06}).  At the same time, the ionizing radiation
may quickly dissociate molecular hydrogen, the only efficient coolant
for low-metallicity gas in such halos, and prevent star formation even
before the gas is completely removed \cite{haiman_etal97}. Even if the
molecular hydrogen is not dissociated, cooling rate in halos with
virial temperature\footnote{The virial virial temperature  $T_{\rm vir}$ 
is related to the virial mass by  $kT_{\rm
vir}=\frac{1}{2}\mu m_p G\Mvir/\Rvir$, where isothermal temperature
profile is assumed for simplicity. The virial mass and radius are
related by definition as $M_{\rm vir}=4\pi/3\times \Delta_{\rm
vir}\bar{\rho}R_{\rm vir}^3$.  Assuming $\Delta_{\rm vir}=178$
appropriate for $z\gtrsim 1$ regardless of $\Omega_0$, this gives
$M_{\rm vir}\approx 5.63\times
10^7h^{-1}\Msun(\Omega_0/0.3)^{-1/2}((1+z)/11)^{-3/2}(T_{\rm
vir}/10^4\,\rm K)^{3/2}$.}  $T_{\rm vir}\lesssim 10^4$~K is
considerably lower than in more massive halos \cite{haiman_etal00} and
we can therefore expect the formation of dense gaseous disks and star
formation suppressed in such halos.  Another potential galaxy
formation suppression mechanism is the gas stripping effect of shocks
from galactic outflows and cosmic accretion
\cite{scannapieco_etal01,sigward_etal04}. Finally, even if the gas is
accreted and cools in small-mass halos, it is not guaranteed that it
will form stars if gas does not reach
metallicities and surface densities sufficient for efficient formation
of molecular gas and subsequent star formation
\cite{kravtsov_etal04b,kaufmann_etal07,tassis_etal08,robertson_kravtsov08,gnedin_etal09}.

The combined effect of these processes is likely to leave most of dark
matter halos with masses $\lesssim {\rm few}\times 10^9\ \rm
M_{\odot}$ dark, and could have imprinted a distinct signature on the
properties of the dwarf galaxies that did manage to form stars before
reionization.  In fact, if all these suppressing effects are as
efficient as is usually thought, it is quite remarkable that galaxies
such as the recently discovered ultra-faint dwarfs exist at all.  One
possibility extensively discussed in the literature is that they
managed to accrete a certain amount of gas and form stars before the
universe was reionized
\cite{bullock_etal00,somerville02,benson_etal02,gnedin_kravtsov06,moore_etal06,madau_etal08,koposov_etal09,busha_etal09}.
Direct cosmological simulations do show that dwarf galaxies forming at
$z>6$ bear striking resemblance to the faint dwarf spheroidal galaxies
orbiting the Milky Way
\cite{ricotti_gnedin05,bovill_ricotti09,ricotti09} and their predicted
abundance around the Milky Way is consistent with estimates of the
abundance of the faintest satellites \cite{gnedin_kravtsov06}.
Alternatively, some authors argued \cite{stoehr_etal02,stoehr_etal03,hayashi_etal03,penarrubia_etal08}
that observed dwarf satellite galaxies could be in much more massive subhalos
than was indicated by simple estimates of 
dynamical masses and circular velocities using stellar velocity dispersions. In
this case, the relatively large halo mass could allow an object to
resist the suppressing effects of the UV background.

Cosmological simulations also clearly show that the subhalos found
within virial radii of larger halos at $z=0$ have on average lost
significant amount of mass and have been considerably more massive in
the past \cite{kravtsov_etal04b,nagai_kravtsov05,giocoli_etal08}. The
dramatic loss of mass occurs due to the tidal forces that subhalos
experience as they orbit in the potential of the host. A significant
fraction of the luminous dwarf satellites therefore can be associated
with those subhalos that have been substantially more massive in the
past and hence more resilient against galaxy formation suppressing
processes. Such subhalos could have had a window of opportunity to
form their stellar systems even if the subhalos they are embedded in
today have relatively small mass \cite{kravtsov_etal04b}.

\subsection{Models for luminous satellite population.}
\label{sec:models}

Given the galaxy formation suppression mechanisms and evolutionary
scenarios listed above, the models aiming to explain the substructure
problem can be split into the following broad classes: 1) the {\it
``threshold galaxy formation models''\/} in which luminous satellites
are embedded in the most massive subhalos of CDM halos and their
relatively small number indicates the suppression of galaxy formation
in subhalos of circular velocity smaller than some threshold value
\cite{stoehr_etal02,stoehr_etal03,penarrubia_etal08} and 2) {\it
``selective galaxy formation models''\/} in which only a fraction of
small subhalos of a given {\it current\/} $\Vmax$ and mass host luminous satellites
while the rest remain dark.

In the second class of models the processes determining whether a
subhalo hosts a luminous galaxy can be the reionization epoch
\cite{bullock_etal00,ricotti_gnedin05,moore_etal06,salvadori_etal08,madau_etal08,busha_etal09}:
subhalos that assemble before the intergalactic medium was heated by
ionized radiation become luminous. The observed faint dwarfs can then
be the ``fossils'' of the pre-reionization epoch \cite{ricotti_gnedin05,bovill_ricotti09}. Subhalo may also
form a stellar system if its mass assembly history was favorable for
galaxy formation \cite{kravtsov_etal04b}: namely, luminous subhalos
are those that have had sufficiently large mass during a period of
their evolution to allow them to overcome the star formation
suppression processes.

Several models using a combination of the processes and scenarios
outlined above have been shown to reproduce the gross properties of observed 
population of satellites reasonably well \cite{kravtsov_etal04b,li_etal09,busha_etal09,koposov_etal09}. 
How can we test different classes of models and differentiate between
specific ones? 

First, I think the fact the $\msixs$ mass function for
observed satellites has a different slope compared to simulation
predictions (Fig.~\ref{fig:mf06}) favors the second class of the
selective galaxy formation models, at least for the brighter
``classical'' satellites. Indeed, given what we know about the average
mass loss of subhalos, it is more natural to
associate the observed systems with the halos of the largest mass {\it
prior to accretion\/} \cite{kravtsov_etal04b} rather than with the subhalos with the largest
current masses. Second, the predicted number of the weakly evolving
pre-reionization objects \cite{gnedin_kravtsov06,madau_etal08,koposov_etal09}, the
extended star formation histories of most of the observed dwarf
satellites\footnote{In fact, in terms of star formation the main
difference between the dIrr and dSph galaxies appears to be presence
or lack of star formation in the last 2 billion years before $z=0$
\cite{orban_etal08}.} \cite{grebel_etal03,dolphin_etal05,orban_etal08,tolstoy_etal09}, and the
significant spread in metallicities and certain isotope ratios
\cite{fenner_etal06} indicate that majority of ``classical'' dwarfs
have not formed most of their stars before reionization but have
formed their stars over rather extended period of time ($\sim
10$~Gyr). It is still possible, however, that a sizable fraction
of the ultra-faint dwarfs are the ``pre-reionization fossils''
\cite{gnedin_kravtsov06,read_etal06,bovill_ricotti09,salvadori_ferrara09}, 
if star formation efficiency in these objects is greatly suppressed \cite{koposov_etal09}
compared to that of brighter dwarfs.

Interesting additional clues and constraints on the galaxy formation
 models available for the dwarf satellites of the Milky Way are the
 measurements of the total dynamical mass within their stellar
 extent. Observations show that the total masses within a fixed
 aperture of the observed satellites are remarkably similar despite a
 several order of magnitude span in dwarf satellite luminosities
 \cite{mateo_etal93,mateo98,gilmore_etal07,penarrubia_etal08,strigari_etal08}. For
 example, the range of masses within $0.6$~kpc shown in
 Figure~\ref{fig:mf06} is only an order of magnitude. Furthermore,
 existence of the tight correlation between total dynamical mass
 within the half-light radius, $M(r_{\rm half})$, and the
 corresponding radius $r_{\rm half}$ \cite{walker_etal09} for bright
 dSph galaxies implies a very similar inner dark matter density
 profile of their host halos.  Recently, L. Strigari and collaborators
 \cite{strigari_etal08} have shown that the mass estimated within the
 central 300~pc, $m(<0.3\,{\rm kpc})\equiv\mthree$, for all of the
 dwarfs with kinematic data varies by at most a factor of four, while
 the luminosity of the galaxies varies by more than four orders of
 magnitude.

\begin{figure}[t]
\vspace{-1.75cm}
\centerline{\includegraphics[width=1.1\linewidth]{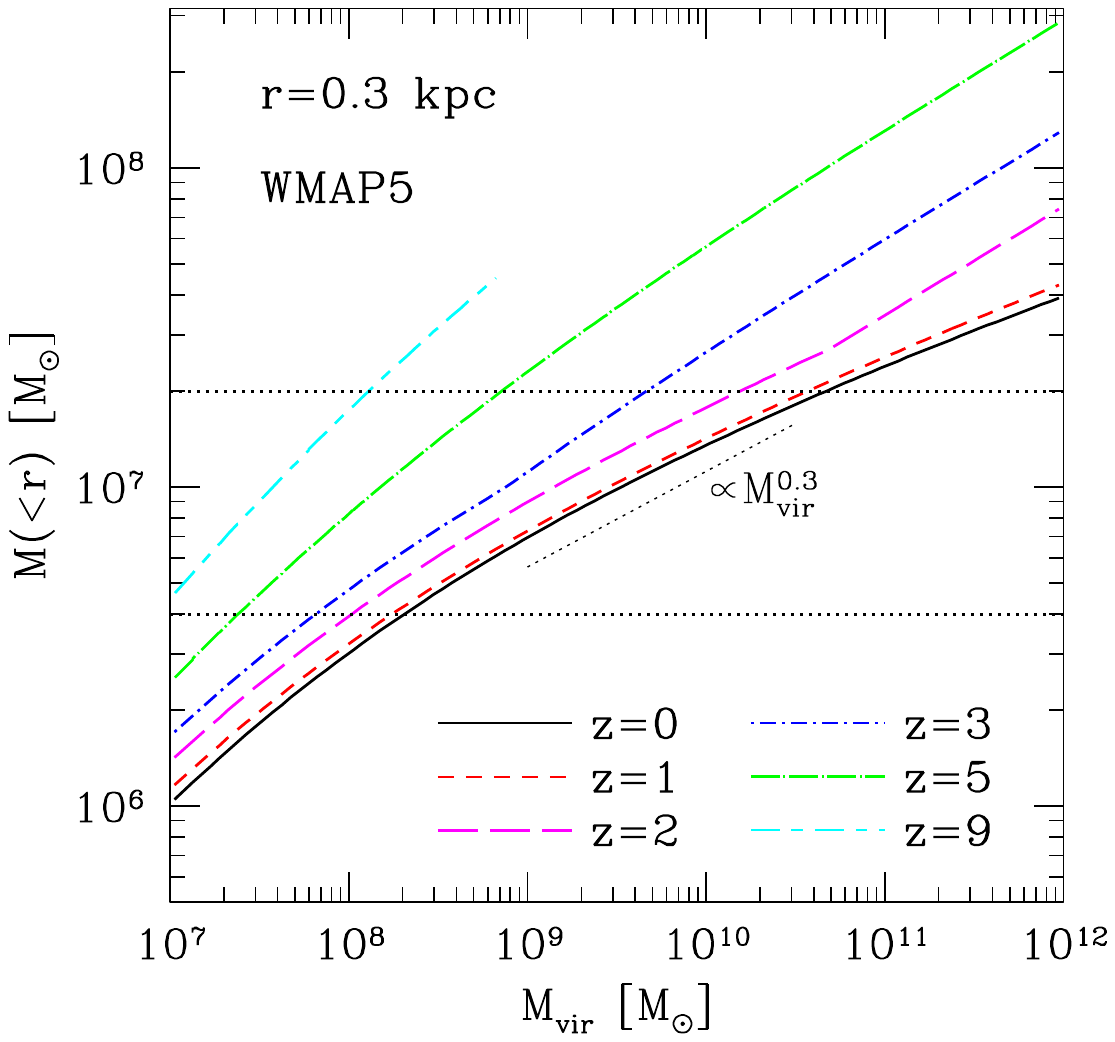}}
\vspace{-2.25cm}
\caption{The mass within central 300 pc vs the total virial mass of an NFW halo
predicted using the concentration mass relation $c(\Mvir)$ at different redshifts 
in the concordance WMAP 5-year best fit cosmology. The horizontal dotted lines
indicate the range of $\mthree$ masses measured for the Milky Way dwarf satellites.}
\label{fig:m03znfw}
\end{figure}

These observational measurements put constraints on the 
range of masses of CDM subhalos that can host observed satellites. 
To estimate this range, we should first note 
that for the CDM halos described by the NFW profile \cite{nfw97} with concentration $c\equiv \Rvir/r_s$ 
(where $r_s$ is the scale radius -- the radius at which the density profile has logarithmic
slope of $-2$) the dependence 
of the mass within a fixed small radius $x\equiv r/\Rvir$ on the total virial mass is
\begin{equation}
m(<x)=\Mvir\frac{f(cx)}{f(c)},\ \rm where 
\end{equation}
\begin{equation}
f(x)\equiv \ln(1+x) - \frac{x}{1+x}
\end{equation}
and is quite weak for $r=300$~pc: $\mthree\propto\Mvir^{0.3-0.35}$, as
shown in Figure~\ref{fig:m03znfw}. Indeed, even for a halo with the
Milky Way mass at $z=0$ we expect $\mthrees\approx 4\times
10^7\,\Msun$, a value not too different from those measured for the
nearby dwarf spheroidals. Physically, the weak dependence of the central mass on the
total mass of the halo reflects the fact that central regions of halos
form very early by mergers of small-mass halos. Given that the rms amplitude 
of density perturbations on small scales is a weak function of scale, the
central regions of halos of different mass form at a similar range of redshifts
and thus have similar central densities reflecting the density of the universe 
when the inner region was assembled. At earlier epochs the dependence is stronger
because 300 pc represents a larger fraction of the virial radius of halos. 

Note that the relation plotted in Figure~\ref{fig:m03znfw} is for
isolated halos unaffected by tidal stripping. Taking into account
effects of tidal stripping results in even flatter relation
\cite{maccio_etal09}: $\mthree\propto\Mvir^{0.25}$, which also has a
lower normalization (smaller $\mthree$ for a given $\Mvir$. This is
likely due to a combination of two effects: 1) the halos of larger
mass have lower concentrations and thus can be stripped more
efficiently and 2) the halos of larger mass can sink to smaller radii
after they accrete and experience relatively more tidal stripping.
Overall, the effect of stripping on $\mthree$ appears to be substantial
and cannot be neglected. 

Finally, figure~\ref{fig:m03znfw} shows that the virial mass range
corresponding to a given range of $\mthree$ is quite different for
halos that form at $z>2$ compared to those that form at later epochs due
to the rapidly evolving NFW concentration concentration for a fixed halo mass \cite{bullock_etal01}.
The mass $\mthree$ can therefore be only interpreted in the context of
a model for subhalo evolutionary histories.
 
Several recent studies have used such models to show that the nearly
constant central mass of the satellite halos is their natural outcome
\cite{li_etal09,maccio_etal09,koposov_etal09}. This outcome can be 
understood as a combination of the weakness of the $\mthree-\Mvir$ 
correlation and the fact that in the galaxy formation models galaxy
luminosity $L$ {\it must} be a nonlinear function of $\Mvir$ in order
to produce a faint-end slope of the galaxy luminosity function
much steeper than the slope of the small-mass tail of the halo mass
function. 

For example, if the faint-end slope of the luminosity function is
$\xi$ (i.e., $dn(L)/dL\propto L^{\xi}$) and the slope of the halo mass
function at small mass end is $\zeta$ ($dn(M)/dM\propto M^{\zeta}$)
and we assume for simplicity a one-to-one monotonic matching between
galaxies and halos $n(>L)=n(>M)$ (see
\cite{conroy_etal06,conroy_wechsler09,tinker_conroy09} for the
detailed justification for such assumption), the implied slope of the
$L-\Mvir$ relation is $\beta=(1+\zeta)/(1+\xi)$, which for the
fiducial values of $\zeta\approx -2$ and $\xi\approx -1.2$ gives
$\beta\approx 5$.  In semi-analytic models, such a steep nonlinear
$L-\Mvir$ relation is usually assumed to be set by either suppression
of gas accretion due to UV heating or by gas blowout due to SN
feedback, e.g., \cite{benson_etal02}). If I instead assume the
faint-end slope of $\xi\approx 1.5-1.6$ as suggested by recent
measurement of \cite{baldry_etal08}, the relation is shallower but is
still nonlinear: $\beta\approx 2$.

Regardless of the specific processes producing the nonlinear
luminosity--mass relation, for a relation of the form $L\propto
\Mvir^{\beta}$ we have $L\propto \mthree^{\gamma}$, where $\gamma\approx
\beta/0.25$, using the $\mthree-\Mvir$ relation above taking into
account effects of tidal stripping. To account for a smaller than a
factor of four spread in central masses for approximately four orders
of magnitude spread in luminosity one needs $\gamma\approx 6-8$ or
$\beta\approx 2-4$, the values not too different from the estimate above.
Thus, {\it the weak correlation of the $\mthree$ 
and luminosity will be the natural outcome of any CDM-based galaxy formation model which 
reproduces the slope of the faint end of the galaxy luminosity function.} 

Within the framework I just described, the slope of the $\mthree-L$
correlation depends on the slope of the $L-M_{\rm vir,acc}$
correlation $\beta$.  The models published so far
\cite{li_etal09,maccio_etal09,koposov_etal09}, as well as the simple
model above with the slope $\beta\approx 2-3$, the slope of the
$L-\mthree$ relation is shallow but is nevertheless not
zero. Constraining this slope with future observations will tighten
constraints on the galaxy formation models and will tell us more about
the $L-\Mvir$ correlation if such exists.

\begin{figure}[t]
\vspace{-5.25cm}
\centerline{\includegraphics[width=1.3\linewidth]{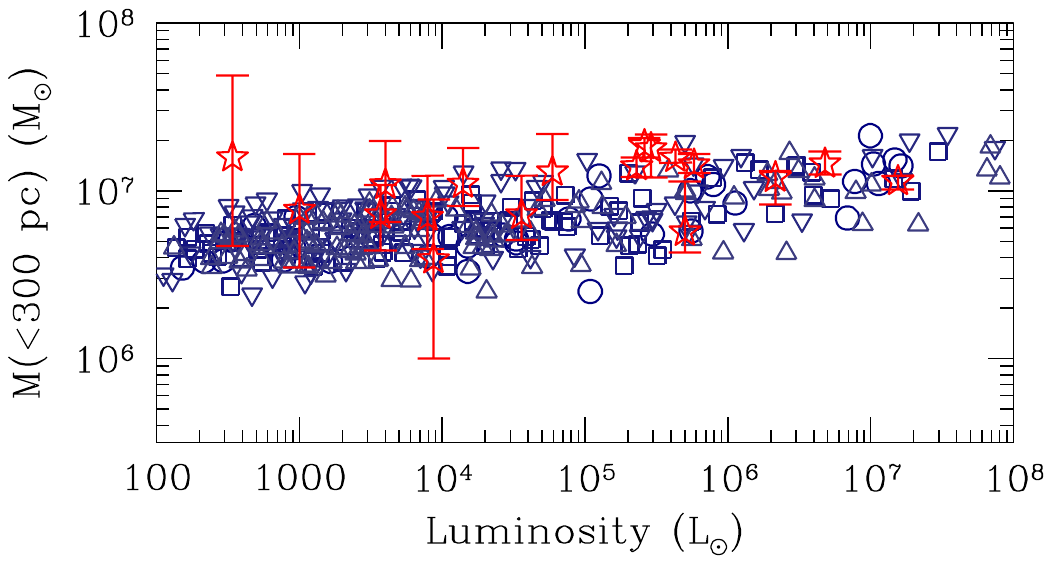}}
\vspace{-3.3cm}
\caption{The mass within the central 300 pc vs luminosity for the dwarf satellites
of the Milky Way (stars with error bars, see \protect\cite{strigari_etal08}). The 
open symbols of different types show the expected relation for subhalos in three
different Milky Way-sized halos formed in the simulations of the
concordance $\Lambda$CDM cosmology  if the luminosity 
of the subhalos is related to their virial mass at accretion epoch as  $L=5\times 10^3\Lsun(M_{\rm vir,acc}/10^9\Msun)^{2.5}$ (see text for discussion).  }
\label{fig:m03l}
\end{figure}

To illustrate the points just made, Figure~\ref{fig:m03l} shows the
$\mthree-L$ relation for the observed nearby dwarfs
\cite{strigari_etal08} and subhalos\footnote{The simulations used here
do not reliably resolve the mass within 300 pc. Therefore, in order to
calculate the mass I have used the mass and concentration of each
subhalo at the epoch when it was accreted and computed evolution of
their density profile given the mass loss they experienced by the
present epoch, as measured in cosmological simulations, and using results of controlled high-resolution
simulations of tidal evolution \cite{penarrubia_etal08b}, which predict the
evolution of the NFW concentration of halos as a function of tidal mass loss. The mass
$\mthree$ was then computed from the evolved density profile.} found
within $300h^{-1}$~kpc around three different MW-sized halos formed in
the concordance cosmological model (see
\protect\cite{kravtsov_etal04b,gnedin_kravtsov06} for simulation
details). To assign luminosity to a given subhalo I follow the 
logic of the model presented in \cite{kravtsov_etal04b}, which 
posits that the brightest observed satellites should correspond
to the subhalos which have the largest mass before they were 
accreted. In this model the luminosity of stellar systems should positively correlate
with the mass of its host subhalo {\it before it was accreted} onto the
MW progenitor, $M_{\rm vir,acc}$:
\begin{equation}
L_V=5\times 10^3\Lsun\left(\frac{M_{\rm vir,acc}}{10^9\Msun}\right)^{2.5}.
\label{eq:lmacc}
\end{equation}
The power law form of the relation is motivated by the approximately
power law form of the galaxy luminosity and halo mass functions at
faint luminosities and small masses. The actual parameters were chosen
such that luminosities of the most massive subhalos roughly match the
luminosities of the most massive satellites, such as the SMC and
LMC\footnote{The figure does not show the most massive subhalos which
would correspond to the systems such as the Large Magellanic Clouds,
which have luminosities $L>10^8\Lsun$ outside the range shown in the
figure. This is justified because observational points shown in the
figure include only the fainter dwarf spheroidal galaxies.} ($L\sim
10^8-10^9\Lsun$). After all, the first order of business for all the
models of satellite population is to reproduce the abundance and
luminosities of the most massive ($\Vmax\gtrsim 40$~km/s) satellites.
The slope of the relation in eq.~\ref{eq:lmacc} was set to reproduce
the range of observed satellite luminosities and flatness of the
$\mthree-L$ relation. Note that this model does not assume any threshold
for formation of galaxies. It simply implies that the efficiency with which
baryons are converted into stars, $f_{\ast}=M_{\ast}/\Mvir$, steadily decreases 
with decreasing $\Mvir$ at the rate given by eq.~\ref{eq:lmacc}.

Figure~\ref{fig:m03l} shows that the model with parameters of eq.~\ref{eq:lmacc} is in agreement with
observed measurements of the $\mthree-L_V$ relation. The results will not change drastically if a 
somewhat steeper ($\approx 3-3.5$) slope is assumed.\footnote{The model of
equation~\ref{eq:lmacc} is similar
to the model 1B in the recent study by Koposov et al. \cite{koposov_etal09}, which
assumes that stellar mass scales as $M_{\ast}=f_{\ast}M_0(M_{\rm vir,acc}/M_0)^{1+\alpha}$, 
These authors find that the model reproduces the luminosity function of satellites for $f_{\ast}\approx 1.7\times 10^{-4}$, $M_0=10^{10}\,\Msun$, and $\alpha=2$, which gives $M_{\ast}=1.7\times 10^3(M_{\rm sat}/10^9\,\Msun)^3$, quite
similar to the relation given by eq.~\ref{eq:lmacc}. Koposov et al. adjust parameters
to match the observed luminosity function and then show that $L-\mthree$ relation is reproduced, 
while I adopted the opposite route here. The key difference between the models is 
that their model assumes a ceiling on the value of $M_{\ast}/M_{\rm vir,acc}$, 
while I assume no such ceiling. Absence of the ceiling on star formation efficiency 
is actually important for 
bright satellites (see Fig.~\ref{fig:fstvm}).}

\begin{figure}[t]
\vspace{-2.cm}
\centerline{\includegraphics[width=1.2\linewidth]{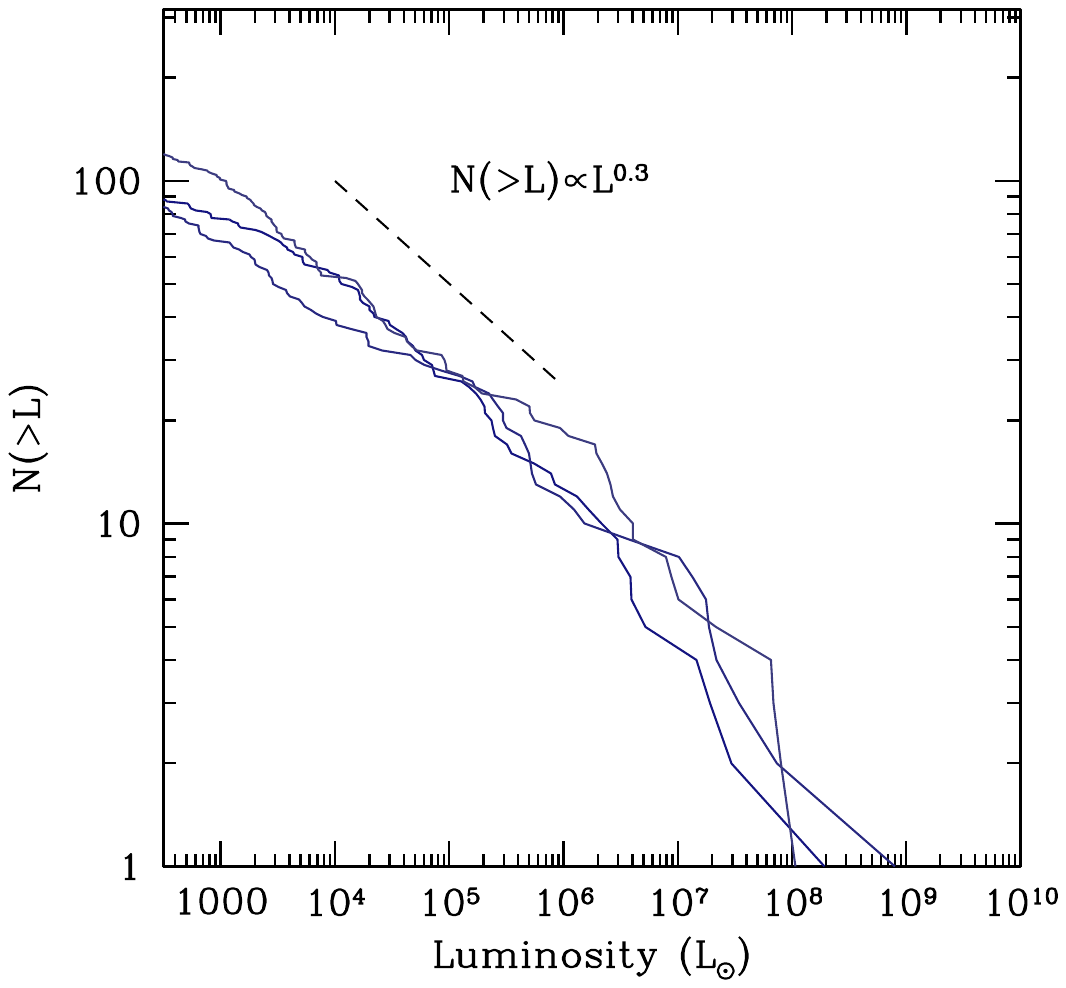}}
\vspace{-3.cm}
\caption{The cumulative luminosity function of subhalos in the three
MW-sized halos shown in Fig.~\ref{fig:m03l} with the simple 
luminosity assignment ansatz of eq.~\ref{eq:lmacc}. The luminosity function
includes all subhalos within 417 kpc from the center of each halo, the
same radius as was used to construct the luminosity function
of the observed satellites shown in Fig.~\ref{fig:lf}. }
\label{fig:lfmodel}
\end{figure}

Having fixed the parameters of the $L-M_{\rm vir,acc}$ relation, we
can then ask the question of whether the luminosity function of
satellites would be reproduced self-consistently by such a model.  We
can use the observed luminosity functions corrected for completeness
for the faintest dwarfs from \cite{tollerud_etal08} (shown in
Fig.~\ref{fig:lf}) to test this. Figure~\ref{fig:lfmodel} shows the
subhalo luminosity functions constructed using subhalos identified
within 417 kpc (the same outer radius used in the construction of the
observed luminosity function by \cite{tollerud_etal08}) in the
simulations and the simple luminosity assignment scheme of
equation~\ref{eq:lmacc}. The luminosity functions in
Fig.~\ref{fig:lfmodel} are in reasonable agreement with observations
within current uncertainties both in their amplitude and slope ($\approx
0.3$).

\begin{figure}[t]
\vspace{-2.cm}
\centerline{\includegraphics[width=1.2\linewidth]{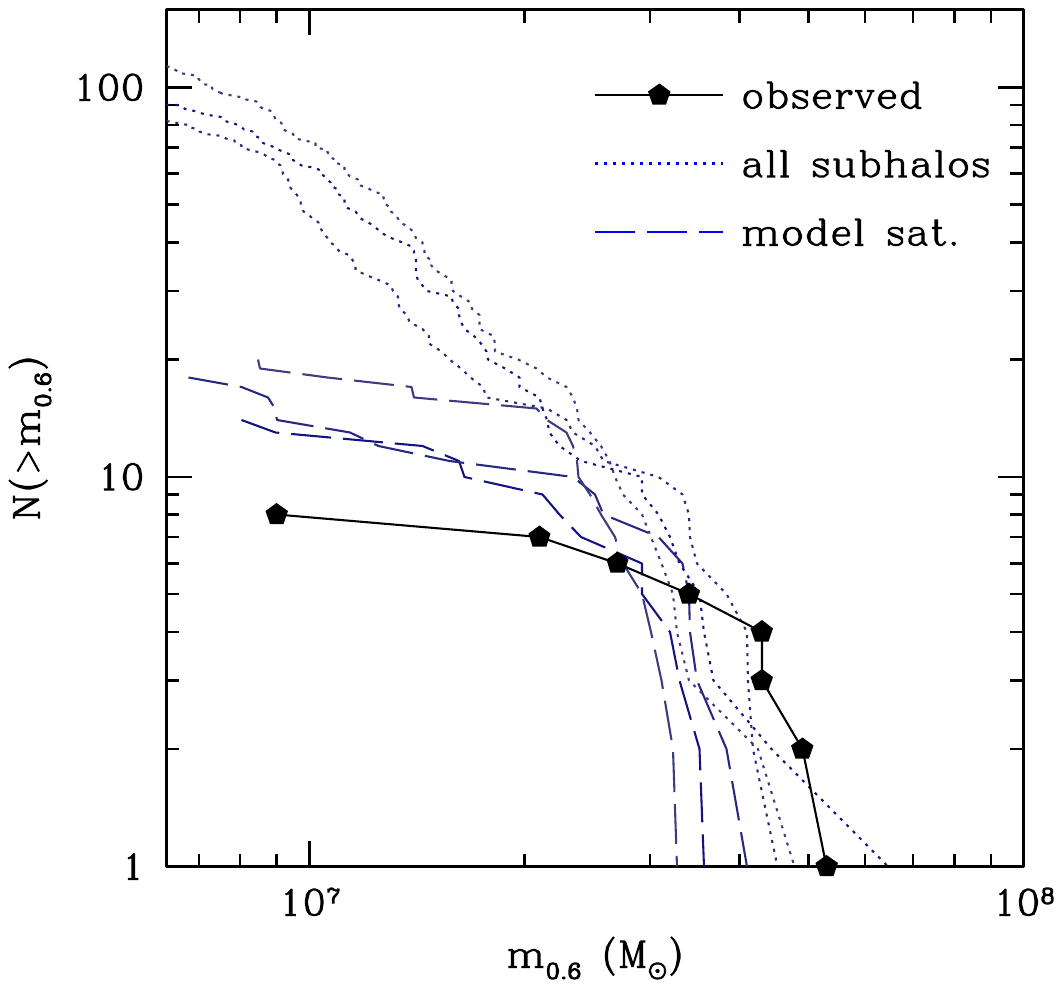}}
\vspace{-3.cm}
\caption{The cumulative $\msixs$ function of subhalos in the $\Lambda$CDM simulations
of three MW halos (dotted lines) and for the observed dSph Milky Way satellites (points, connected
by solid line) Sagittarius
was excluded due to its very large mass errors), as measured by \cite{strigari_etal07}. The
dashed lines show the mass function for subhalos with $L\geq 2.6\times 10^5\,\Lsun$ and $d<270$~kpc
(the same range of luminosities and distances as for the observed satellites) with luminosities
assigned using equation \ref{eq:lmacc}. I have excluded the two most luminous objects from the model
luminous satellites to account for the fact that SMC and LMC are not included in the observational
sample and one additional object to account for non-inclusion of the Sagittarius dwarf in the comparison.}
\label{fig:mf06model}
\end{figure}

Figure~\ref{fig:mf06model} shows comparison of the $\msixs$ mass functions
for the observed ``classical'' dwarf spheroidal satellites of the Milky Way
\cite{strigari_etal07} and predicted mass function for the entire
subhalo population of the MW-sized halos within 270~kpc (the largest
distance of the observed satellite included in the comparison) and the
subhalos with $L>2.6\times 10^5\,\Lsun$ (the smallest dSph luminosity
included in the observed sample) with luminosities assigned using
eq.~\ref{eq:lmacc}. I have excluded the Sagittarius dwarf from this
comparison as its $\msixs$ mass has very large errors
\cite{strigari_etal07}. The number of the predicted luminous
satellites was reduced by three to account for exclusion of
the Sagittarius, SMC, and LMC from the comparison. The figure shows that
the model predicts the range of $\msixs$ quite similar to that
measured for the observed luminous dSphs. The shape of the mass function is also in reasonably good
agreement with the data. Although there are somewhat more predicted
satellites at small masses, this is likely due to the somewhat larger
virial mass of the simulated halos ($\approx 2-3\times
10^{12}\,\Msun$) compared to the mass of the Milky Way ($\approx
10^{12}\,\Msun$). We expect the number of subhalos to scale approximately
linearly with host mass and so the difference in the virial mass of the Milky Way
and simulated halos can account for the difference with observations in Figure~\ref{fig:mf06model}. 
There is some discrepancy at the largest $msixs$ values, 
but it is not clear just how significant the discrepancy is given that
the typical errors on the $\msixs$ measurements for these galaxies are $\approx
20-40\%$. 

\begin{figure}[t]
\vspace{-2.cm}
\centerline{\includegraphics[width=1.2\linewidth]{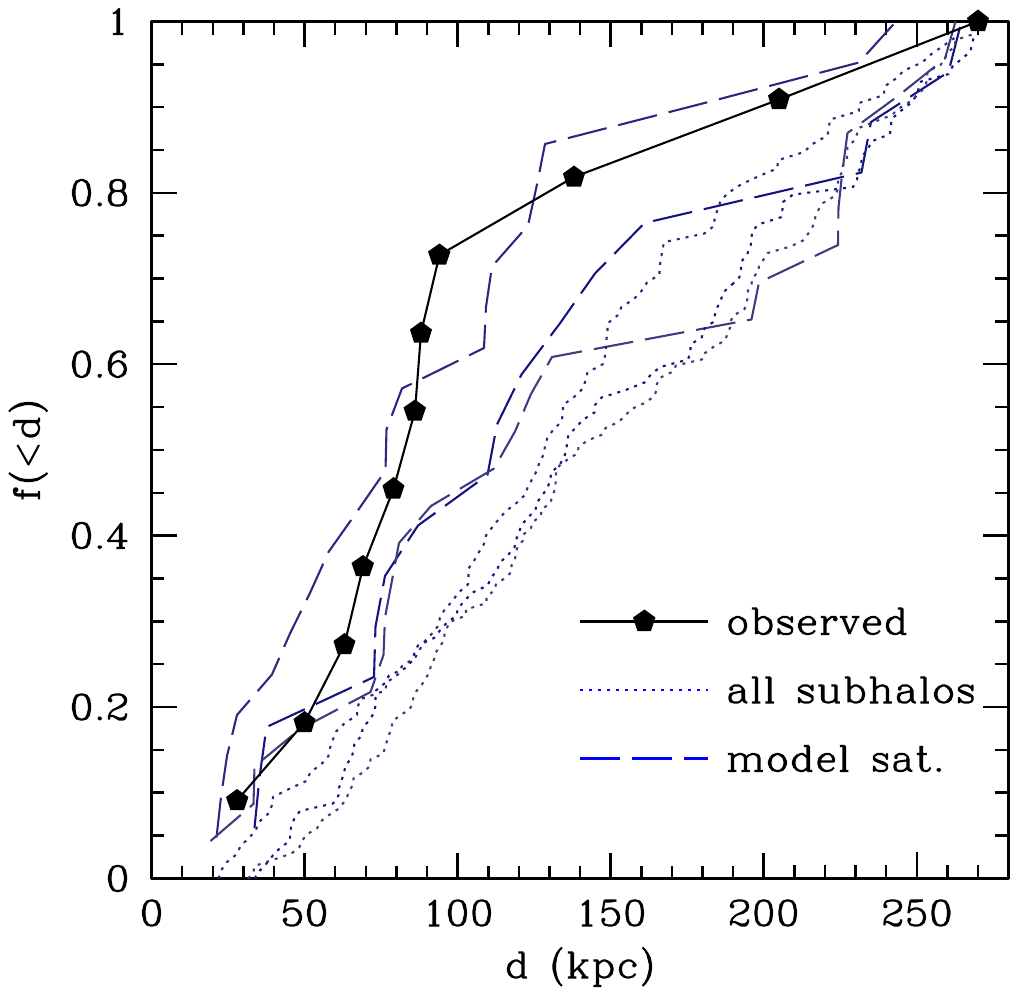}}
\vspace{-3.cm}
\caption{The cumulative radial
distribution of the observed ``classical'' Milky Way satellites (solid points connected 
by the solid line) within
280 kpc and satellites with similar luminosities and within the same
distance from their host halo in the model of eq.~\ref{eq:lmacc} (dashed lines).
The figure also shows the cumulative distribution of all subhalos
selected using their current $\Vmax$ (dotted lines).}
\label{fig:rdmodel}
\end{figure}

\begin{figure}[t]
\vspace{-2.cm}
\centerline{\includegraphics[width=1.2\linewidth]{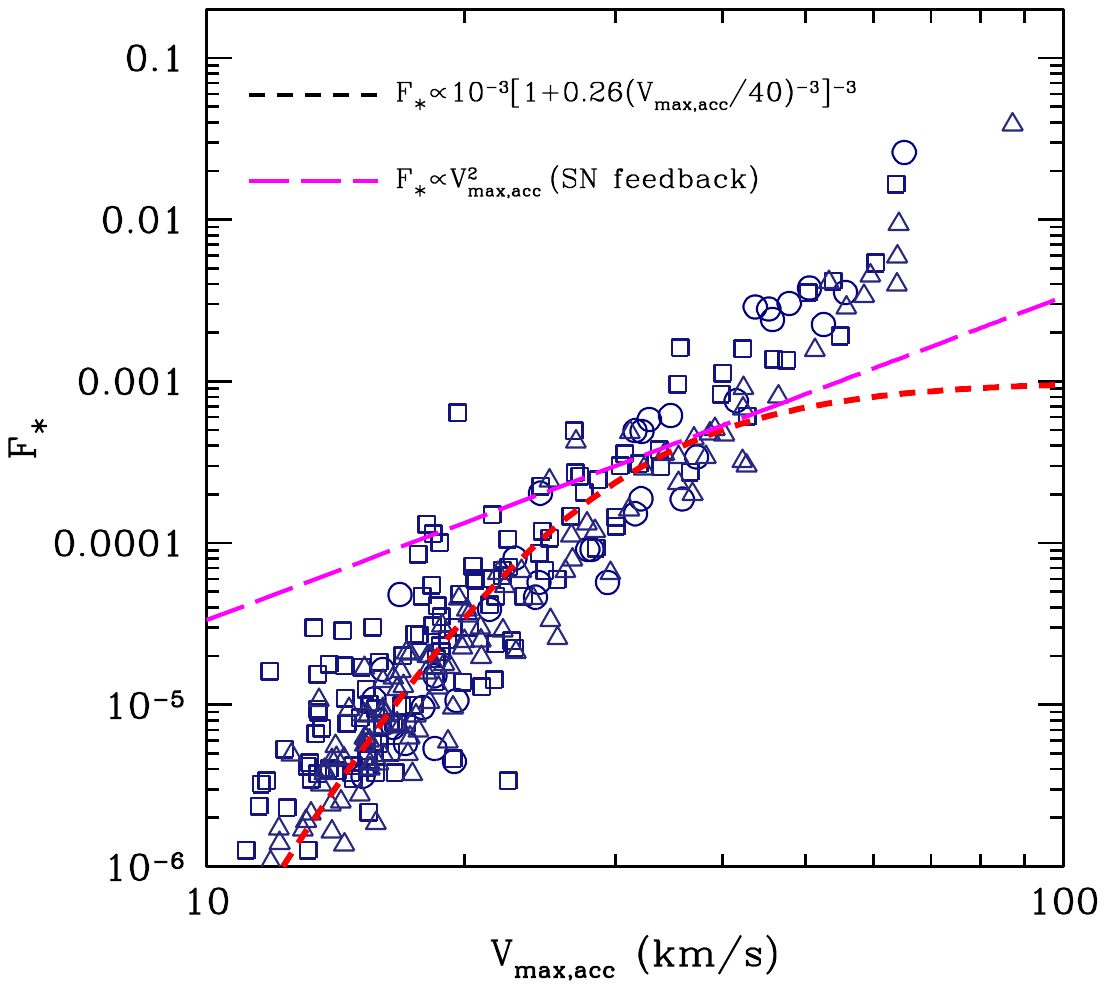}}
\vspace{-3.cm}
\caption{The efficiency of gas conversion into stars, defined as the fraction of 
baryon mass, expected if the halo accreted its universal fraction of baryons, converted
into stars: $F_{\ast}=(\Omega_m/\Omega_b)(M_{\ast}/M_{\rm vir,acc})$. The points (as before
different symbols correspond to subhalos in three different simulated host halos) show
the dependence of $F_{\ast}$ on the maximum circular velocity of each subhalo
at accretion according to luminosity assignment model of eq.~\ref{eq:lmacc}. The short-dashed
line shows the functional form expected if $F_{\ast}$ was controlled by the 
suppression of gas accretion due to UV heating of intergalactic gas (model 3B of \cite{koposov_etal09}),
while long-dashed line shows scaling expected in efficiency was set by supernova feedback (e.g., \cite{dekel_woo03}).}
\label{fig:fstvm}
\end{figure}

Finally, figure~\ref{fig:rdmodel} compares cumulative radial
distribution of the observed ``classical'' Milky Way satellites within
280 kpc and satellites with similar luminosities and within the same
distance from their host halo in the model of eq.~\ref{eq:lmacc}.
The figure also shows the cumulative distribution of all subhalos
selected using their current $\Vmax$. The predicted distribution
of bright luminous satellites is somewhat more radially 
concentrated than the distribution of the $\Vmax$-selected subhalos
and is in reasonable agreement with the observed distribution both in
its median and in the overall shape. 

Thus, the observed $\mthree-L$ relation, the luminosity function, the
$\msixs$ mass function, and the radial distribution of the observed
satellites can all be reproduced simultaneously with such a simple dwarf
galaxy formation scenario. The observational uncertainties in these
statistics are still quite large, which leaves significant freedom in
the parameters of eq.~\ref{eq:lmacc} and in its functional
form.\footnote{It is quite possible that relation between luminosity
and mass is more complicated than eq.~\ref{eq:lmacc}. For example, the
normalization of the $L-M_{\rm vir,acc}$ relation can evolve with
redshift.because luminosity may be determined both by the mass of the
halo at the accretion epoch and by the period of time before its
accretion during which it was sufficiently massive to withstand star
formation suppressing processes. Such redshift dependence would be an
extra parameter which would generate scatter in the $L-M_{\rm
vir,acc}$ relation. } It is also possible that all of these statistics
may be reproduced in a drastically different scenario. Nevertheless,
the success of such simple model is encouraging and it is interesting
to discuss its potential implications.

First of all, the equation~\ref{eq:lmacc} implies that all of the observed Milky
Way dSph satellites had virial masses $M_{\rm vir,acc}\gtrsim 5\times 10^8\,\Msun$
when they were accreted and these masses may span the range up to
$\sim 5\times 10^{10}\,\Msun$ (the actual range depends sensitively on
the slope of the $L-M_{\rm vir,acc}$ relation).  This shows 
that progenitors of the observed satellites could have had a wide range of virial
masses, even though the range of their $\mthree$ and $\msixs$ masses is narrow.

An interesting implication of the value of lowest mass of the range of
masses above is that whatever gas the small-mass halos ($\Mvir\lesssim
5\times 10^8\,\Msun$) are able to accrete, it should remain largely
unused for star formation, and of course would not be blown away by
supernovae (given that the model implies that such objects should have no
stars or supernovae).  If some of this gas is neutral, it can contribute to HI
absorption lines in the spectra of quasars and distant galaxies. If
this gas is enriched, it can also produce absorption lines of heavier
elements. At lower redshifts, the neutral gas in the otherwise
starless or very faint halos could manifest itself in the form of the
High Velocity Clouds (HVCs) abundant in the Local Group
\cite{braun_burton99,deheij_etal02,thilker_etal04,grossi_etal08} and
around other galaxies.

Second, as I noted above the slope of the $L-\Mvir$ relation required
to explain the weak dependence of $\mthree$ on luminosity is not
surprising, given what we know about the faint-end slope of galaxy
luminosity function and what we expect about the slope of the mass
function of their host halos in CDM scenario
\cite{vandenbosch_etal07}. The implied {\it normalization\/} of the
$L-\Mvir$ relation, however, is quite interesting. For example, it
indicates that halo of $M_{\rm vir,acc}=10^{10}\,\Msun$ should have
luminosity of $L_V=1.6\times 10^6\,\Lsun$. Converting it to stellar
mass assuming $M_{\ast}/L_V=1$ (appropriate for old populations,
e.g. \cite{martin_etal08}) gives $M_{\ast}=1.6\times 10^6\,\Msun$.
Results of cosmological simulations with UV heating of gas show that
halos of $M\sim 10^{10}\,\Msun$ should have been able to accrete
almost all of their universal share of baryons,
$M_b=(\Omega_b/\Omega_m)M_{\rm vir,acc}\approx 1.7\times 10^9\,\Msun$
(assuming $\Omega_b/\Omega_m\approx 0.17$ suggested by the {\sl
WMAP\/} measurements \cite{dunkley_etal09}), even in the presence of
realistic UV heating \cite{hoeft_etal06,okamoto_etal08}.  The derived
stellar mass thus implies that only $F_{\ast}\equiv M_{\ast}/M_{\rm
vir,acc}\times (\Omega_m/\Omega_b)\approx 0.001$ (i.e., $0.1\%$) of
the expected baryon mass was converted into stars in such objects.
Such small efficiency $F_{\ast}$ for systems accretion onto which is
not suppressed by the UV heating implies that {\it star formation is
dramatically suppressed by some other mechanism.}

In fact, the implied efficiency of baryon conversion into the stars,
$F_{\ast}$, in this model is a steep, power law function of mass and
circular velocity ($F_{\ast}\approx V_{\rm max,acc}^{8\div 9}$), as
shown in figure~\ref{fig:fstvm}. The figure also shows the functional
form one would have expected if the dependence of the efficiency
$F_{\ast}$ on circular velocity was determined by the fraction of gas
halos are able to accrete in the presence of the UV radiation. This
functional form is almost identical to the fiducial model 3B of
Koposov et al. \cite{koposov_etal09} (I used slightly larger value for
critical velocity because I use $\Vmax$ rather than the virial
circular velocity used by these authors).  The two models have similar
behavior at $V_{\rm max,acc}\lesssim 30$~km/s, but the UV heating
model asymptotes to a fixed value of $F_{\ast}=10^{-3}$ for more
massive systems. This model would therefore underpredict luminosities
of the most massive satellites in MW-sized halos, which was noticed by
Koposov et al. in its failure to reproduce the bright end of the
satellite luminosity function. Moreover, the estimates of the
efficiency $F_{\ast}$ for more luminous galaxies, such as the Milky
Way, are in the range $F_{\ast}\sim 0.05-0.2$
\cite{pizagno_etal05,mandelbaum_etal06,conroy_etal07,klypin_etal09},
which would lie roughly on the continuation of the relation in
Fig.~\ref{fig:fstvm} to larger circular velocities.\footnote{The
$F_{\ast}-\Vmax$ relation could plausibly flatten at larger $\Vmax$,
as expected from the halo modeling of the galaxy population (e.g.,
\cite{vandenbosch_etal07})}.

These considerations indicate a very interesting possibility that
while the UV heating can mediate accretion of gas into very small mass
halos, the efficiency with which the accreted gas is converted into
stars in the known luminous galaxies is dramatically suppressed by
some other {\it mass-dependent\/} mechanism. This strong suppression
operates not only for halos in which gas accretion is suppressed but
for halos of larger masses as well. Note that this suppression mechanism
is unlikely to be due to blowout of gas by supernovae, which is expected
to give $F_{\ast}\propto \Vmax^2$ (e.g., \cite{dekel_woo03}), a much 
shallower relation than the scaling in Fig.~\ref{fig:fstvm}.

While discussion of the nature of this suppression mechanism is
outside the scope of this paper, the rapidly improving observational
data on the satellite population should shed light on the possible
mechanisms. 

The exercise presented in this section illustrates just
how powerful the combination of luminosity function and radial
distribution of satellites, high quality resolved kinematics data and
inferred dynamical constraints on the total mass profile, measurements
of star formation histories and enrichment histories can be in
understanding formation of dwarf galaxies. The rapidly improving
constraints on the mass profiles of the dSph galaxies down to the
smallest luminosities \cite{strigari_etal08,walker_etal09} should
further constrain the range of subhalo masses hosting the observed
satellites and, by inference, the efficiency of star formation in such
small halos. 
 The hints that star
formation efficiency is actually a monotonic function of halo mass
from galaxies such as the Milky Way to the faintest known galaxies,
such as Segue 1, indicates that the physics learned from the
``near-field cosmology'' studies of the nearest dwarfs can potentially give
us important insights into formation of more massive galaxies as well.

\noindent
{\bf Acknowledgements.} I would like to thank Brant Robertson, Anatoly
Klypin, Nick Gnedin, Jorge Pe\~narrubia, Erik Tollerud, James Bullock
for useful comments on the manuscript and stimulating discussions on
the topics related to the subject of this paper.  This work was
partially funded by the NSF grants AST-0507666, and AST-0708154.  The
research was also partially supported by the Kavli Institute for
Cosmological Physics at the University of Chicago through grant NSF
PHY-0551142 and an endowment from the Kavli Foundation. I would like
to thank Kavli Institute for Theoretical Physics (KITP) in Santa
Barbara and organizers of the 2008 KITP workshop ``Back to the Galaxy
II,'' where some of the work presented here was carried out, for
hospitality and wonderful atmosphere. I have made extensive use of the
NASA Astrophysics Data System and arXiv.org preprint server during
writing of this paper.

\bibliography{sat}

\end{document}